\documentclass{article}
\usepackage{epsfig}
\usepackage[hang,nooneline]{subfigure}
\date{\today}

\usepackage{amsmath}
\usepackage{amssymb}

\begin{document}
\title{Supersymmetric $Q$-balls and boson stars in $(d+1)$ dimensions}
\author{
{\bf Betti Hartmann\footnote{b.hartmann@jacobs-university.de} \ \ and \
J\"urgen Riedel\footnote{j.riedel@jacobs-university.de}}\\
School of Engineering and Science, Jacobs University, 28725 Bremen, Germany}

\newcommand{\dd}{\mbox{d}}
\newcommand{\tr}{\mbox{tr}}
\newcommand{\la}{\lambda}
\newcommand{\ka}{\kappa}
\newcommand{\f}{\phi}
\newcommand{\vf}{\varphi}
\newcommand{\F}{\Phi}
\newcommand{\al}{\alpha}
\newcommand{\ga}{\gamma}
\newcommand{\de}{\delta}
\newcommand{\si}{\sigma}
\newcommand{\bomega}{\mbox{\boldmath $\omega$}}
\newcommand{\bsi}{\mbox{\boldmath $\sigma$}}
\newcommand{\bchi}{\mbox{\boldmath $\chi$}}
\newcommand{\bal}{\mbox{\boldmath $\alpha$}}
\newcommand{\bpsi}{\mbox{\boldmath $\psi$}}
\newcommand{\brho}{\mbox{\boldmath $\varrho$}}
\newcommand{\beps}{\mbox{\boldmath $\varepsilon$}}
\newcommand{\bxi}{\mbox{\boldmath $\xi$}}
\newcommand{\bbeta}{\mbox{\boldmath $\beta$}}
\newcommand{\ee}{\end{equation}}
\newcommand{\eea}{\end{eqnarray}}
\newcommand{\be}{\begin{equation}}
\newcommand{\bea}{\begin{eqnarray}}

\newcommand{\ii}{\mbox{i}}
\newcommand{\e}{\mbox{e}}
\newcommand{\pa}{\partial}
\newcommand{\Om}{\Omega}
\newcommand{\vep}{\varepsilon}
\newcommand{\bfph}{{\bf \phi}}
\newcommand{\lm}{\lambda}
\def\theequation{\arabic{equation}}
\renewcommand{\thefootnote}{\fnsymbol{footnote}}
\newcommand{\re}[1]{(\ref{#1})}
\newcommand{\R}{{\rm I \hspace{-0.52ex} R}}
\newcommand{\N}{{\sf N\hspace*{-1.0ex}\rule{0.15ex}%
{1.3ex}\hspace*{1.0ex}}}
\newcommand{\Q}{{\sf Q\hspace*{-1.1ex}\rule{0.15ex}%
{1.5ex}\hspace*{1.1ex}}}
\newcommand{\C}{{\sf C\hspace*{-0.9ex}\rule{0.15ex}%
{1.3ex}\hspace*{0.9ex}}}
\newcommand{\eins}{1\hspace{-0.56ex}{\rm I}}
\renewcommand{\thefootnote}{\arabic{footnote}}

\maketitle

\bigskip

\begin{abstract}
We construct
supersymmetric $Q$-balls and boson stars in $(d+1)$ dimensions.
These non-topological solitons are solutions of a scalar field model with global $U(1)$ symmetry
and a scalar field potential that appears 
in gauge-mediated supersymmetry (SUSY) breaking in the minimal supersymmetric
extension of the Standard Model (MSSM). We are interested in both the asymptotically flat as well as in
the asymptotically Anti-de Sitter (AdS) solutions. In particular, we show that for  our choice of the potential
gravitating, asymptotically flat boson stars exist in $(2+1)$ dimensions. We observe that the behaviour
of the mass and charge of the asymptotically flat solutions at the approach of the maximal frequency depends strongly
on the number of spatial dimensions. In particular, we find that in the ``thick-wall limit'' $Q$-balls are always unstable 
in flat space-time, but that they can become stable in AdS. 
Moreover, for the asymptotically AdS solutions the model on the conformal boundary
can be interpreted as describing $d$-dimensional condensates of scalar glueballs. 
\end{abstract}

\section{Introduction}
A number of non-linear field theories possess solitonic-like solutions. These have broad applications in many branches
of physics and constitute localized, globally regular structures with finite energy. Topological solitons \cite{ms}
possess a conserved topological charge that is connected to the existence of non-contractible loops in the theory.
Non-topological solitons \cite{fls,lp} on the other hand appear in models with symmetries and possess a locally conserved
Noether current and a globally conserved Noether charge. An example of such a non-topological soliton is the $Q$-ball \cite{coleman}
and its generalization in curved space-time, the boson star \cite{kaup,misch,flp,jetzler,new1,new2}. These are solutions
of models with self-interacting complex scalar fields and the conserved Noether charge 
$Q$ is then related to the global phase invariance of the theory and is directly proportional
to the frequency of the harmonic time-dependence. $Q$ can e.g. be interpreted as particle number \cite{fls}. 
As such, these solutions have been constructed in $(3+1)$-dimensional models with 
non-renormalizable $\Phi^6$-potential \cite{vw,kk1,kk2}, but also in supersymmetric extensions to the
Standard Model (SM) \cite{kusenko}. 
In the latter case, several scalar fields
interact via complicated potentials. It was shown that cubic interaction terms that result from
Yukawa couplings in the superpotential and supersymmetry (SUSY) breaking terms lead to the existence of $Q$-balls
with non-vanishing baryon or lepton number or electric charge. These supersymmetric
$Q$-balls have been considered as possible candidates for baryonic dark matter 
\cite{dm} and their astrophysical implications have been discussed \cite{implications}.
In \cite{cr}, these objects have been constructed numerically using 
the exact form of a scalar potential that results from gauge-mediated SUSY breaking. However, this
potential is non-differentiable at the SUSY breaking scale.
In \cite{ct} a differentiable approximation of this potential was suggested and the
properties of the corresponding $Q$-balls have been investigated. Most models of a quantum theory of gravity
need more than $(3+1)$ dimensions and as such, it is surely of interest to investigate the properties of
soliton solutions in higher dimensions. The first study of $Q$-balls in higher dimensional space-time has been
done in \cite{multamaki_vilja,prikas1}. In \cite{multamaki_vilja} a mixture of analytical
and numerical tools was used, while in \cite{prikas1} only a linearized version of the Lagrangian and equations
of motion depending only in zeroth order on the ratio between the typical energy scale and the Planck mass
has been used. In this case, an analytical solution can be given, however, the model does not
capture the non-linear phenomena such as e.g. the behaviour of the mass and charge at the maximal frequency. 
A similar study has been done in \cite{prikas2} for $(2+1)$ dimensions.
$Q$-balls and boson star solutions of the full system of coupled non-linear equations 
in $(4+1)$-dimensional asymptotically flat
space-time have been investigated in \cite{hartmann_kleihaus_kunz_list}. Interestingly, it was found
that the behaviour of the mass and charge at the approach of the maximal possible frequency is different for
$d=4$ as compared to $d=3$. This was related to a scaling behaviour of the solutions at this critical approach
and different dimensions of the spatial integrals.
Spinning generalisations of these solutions can also be constructed 
\cite{vw,kk1,kk2,hartmann_kleihaus_kunz_list,bh,radu_aste}. These solutions possess a quantised angular momentum
that is an integer multiple of the Noether charge.

Topological and non-topological solitons in Anti-de Sitter (AdS) space-time have been investigated 
intensively recently. The interest in these
objects is related to the AdS/CFT correspondence \cite{ggdual,adscft} which states that a gravity theory in a $d$-dimensional
Anti-de Sitter (AdS) space--time is equivalent to a Conformal Field Theory (CFT) on the $(d-1)$-dimensional boundary of AdS.
Interestingly, this is a weak-strong coupling duality that can be used to describe strongly coupled
Quantum Field Theories with the help of weakly coupled gravity theories. This has been applied to a modeling
of high temperature superconductivity with the help of classical black hole and soliton solutions in AdS
 \cite{gubser,hhh,reviews}. The basic models use a scalar field coupled to a $U(1)$ gauge field and the observation
that close to the horizon of the black hole the effective mass of the scalar field can become
negative with masses below the Breitenlohner--Freedman bound \cite{bf} such that the scalar
field becomes unstable and possesses a non--vanishing value on and close to the horizon
of the black hole. When computing the conductivities it turns out that the formation of a scalar field
on a charged black hole corresponds to a phase transition from a conductor to a superconductor.
However, insulator/superconductor phase transitions also play an important role in high temperature superconductivity
and as such models including solitons have been suggested that describe this phenomenon 
\cite{nrt,horowitz,hartmann_brihaye2}.
The AdS soliton is related to the black hole by a double Wick rotation with one of the coordinates 
compactified to a circle and has originally been suggested to describe a confining vacuum 
in the dual gauge theory \cite{witten2,horowitz_myers} since it possesses a mass gap.
For solutions with Ricci-flat horizons 
there is a phase transition between the AdS black hole and the AdS soliton \cite{ssw} which was interpreted as
a confining/deconfining phase transition in the dual gauge theory.
Note that this is different for  black holes in global AdS where the black hole decays to global AdS space-time when
lowering the temperature \cite{hawking_page}. 

In the limit of vanishing gauge coupling the soliton
solutions correspond to planar boson stars in AdS space-time. Since the scalar field
is uncharged the interpretation in terms of insulators/superconductors is difficult in this
case. However, since the AdS/CFT correspondence connects strongly coupled
CFTs to weakly coupled gravity theories the prototype example of a strongly coupled
field theory comes to mind - Quantum Chromodynamics (QCD). As such the planar boson stars in AdS
have been interpreted
as Bose-Einstein condensates of glueballs. Glueballs are color-neutral bound states of gluons predicted by QCD
and the scalar glueball (which is also the lightest possible glueball) is predicted to have a mass of 1-2 GeV (see e.g.
\cite{glueballs_experiment} for an overview on experimental results). Since these glueballs
appear due to non-linear interactions and as such cannot be described by a perturbative approach
it is very difficult to make predictions within the framework of Quantum Field Theory.  
However, holographic methods
have been applied to make predictions about glueball spectra (see e.g. \cite{physics_glueballs} and 
reference therein).

Non-spinning boson stars in $(d+1)$-dimensional AdS space-time have been studied before using a massive
scalar field without self-interaction \cite{radu} and in $(3+1)$ dimensions with an exponential self-interaction
potential \cite{hartmann_riedel}. Spinning solutions in $(2+1)$ and $(3+1)$ dimensions
have been constructed in \cite{radu_aste} and \cite{radu_subagyo}, respectively. 

In this paper, we are interested in $Q$-balls and boson stars in both asymptotically flat as well as
asymptotically AdS space-time with $(d+1)$ dimensions. We use an exponential scalar field potential already
employed in \cite{hartmann_riedel}.
We will consider first the asymptotically flat case generalising some of
the results obtained in \cite{hartmann_kleihaus_kunz_list} to higher dimensions and then also consider asymptotically
AdS solutions. 

Our paper is organised as follows: In Section 2 we give the model, equations of motion and boundary conditions.
In Section 3, we present our numerical results and conclude in Section 4. The Appendix 1 and 2 contain results
on the generalisation of an exact
solution first found in \cite{radu_subagyo} to $(d+1)$ dimensions and
on the existence of asymptotically flat boson stars in $(2+1)$ dimensions, respectively.  

\section{The model}

In the following we will study non-spinning $Q$-balls and boson stars in a $(d+1)$-dimensional Anti-de Sitter
(AdS) space time. 
The action $S$ reads
\begin{equation}
\label{action}
 S=\int \sqrt{-g} d^{d+1} x \left( \frac{R-2\Lambda}{16\pi G_{d+1}} + {\cal L}_{m}\right) + \frac{1}{8\pi G_{d+1}} 
\int d^{d} x \sqrt{-h} K
\end{equation}
where $R$ is the Ricci scalar, $G_{d+1}$ denotes the $(d+1)$-dimensional Newton's constant, 
$\Lambda$ is the negative cosmological constant related to the AdS radius $\ell$ by
$\Lambda=-d(d-1)/(2\ell^2)$. The second term on the right hand side of (\ref{action}) 
is the Gibbons-Hawking surface term \cite{gibbons_hawking} with $h$ the induced metric 
and $K$ the trace of the extrinsic curvature on the AdS boundary. ${\cal L}_{m}$ is
the matter Lagrangian given by
\begin{equation}
\label{lag}
 {\cal L}_{m}=-\partial_{M} \Phi \partial^{M} \Phi^*
 - U(\vert\Phi\vert) \ \ , \ \ M=0,1,....,d \ ,
\end{equation}
where $\Phi$ denotes a complex scalar field and we choose the metric to have mainly positive signature.
$U(\vert\Phi\vert)$ is the potential 
\begin{equation}
\label{potential}
 U(\vert\Phi\vert)=m^2\eta_{\rm susy}^2 \left(1-\exp\left(-\frac{\vert\Phi\vert^2}{\eta_{\rm susy}^2}\right)\right) \ .
\end{equation}
This potential is motivated by supersymmetric extensions of the Standard Model \cite{cr,ct}.
Here $\eta_{\rm susy}$ is a parameter such that $\eta_{\rm susy}^{2/(d-1)}$ corresponds to the energy scale 
below which supersymmetry is broken, while $m$ denotes the scalar boson mass.

The coupled system of ordinary differential equations is then given by the Einstein
equations
\begin{equation}
\label{einstein}
 G_{MN}+\Lambda g_{MN}=8\pi G_{d+1} T_{MN} \ \ , \ \ M,N=0,1,..,d
\end{equation}
with the energy-momentum tensor
\begin{eqnarray}
\label{em}
T_{MN}&=& g_{MN} {\cal L} - 2\frac{\partial {\cal L}}{\partial g^{MN}}\nonumber\\
&=& -g_{MN} \left[\frac{1}{2} g^{KL} 
\left(\partial_{K} \Phi^* \partial_{L} \Phi +
\partial_{L} \Phi^* \partial_{K} \Phi\right) + U(\Phi)\right] +
\partial_{M} \Phi^* \partial_{N} \Phi + \partial_{N}\Phi^* \partial_{M} \Phi 
\end{eqnarray}
and the Klein-Gordon equation
\begin{equation}
\label{KG}
 \left(\square - \frac{\partial U}{\partial \vert\Phi\vert^2} \right)\Phi=0 \ \   \ \ .
\end{equation}
The matter Lagrangian ${\cal L}_{m}$ (\ref{lag}) is invariant under the global U(1) transformation
\begin{equation}
 \Phi \rightarrow \Phi e^{i\chi} \ \ \  .
\end{equation}
As such the locally conserved Noether
current $j^{M}$, $M=0,1,..,d$ associated to this symmetry is given by
\begin{equation}
j^{M}
 = -\frac{i}{2} \left(\Phi^* \partial^{M} \Phi - \Phi \partial^{M} \Phi^*\right) \  \ {\rm with} \ \ \
j^{M}_{; M}=0  \ .
\end{equation}
The globally conserved Noether charge $Q$ of the system then reads
\begin{equation}
 Q= -\int d^{d} x \sqrt{-g} j^0  \  .
\end{equation}

\subsection{Ansatz and Equations}
For the metric we use the following Ansatz in spherical Schwarzschild-like coordinates
\begin{equation}
 ds^2 = - A^2(r)N(r)dt^2 + \frac{1}{N(r)}dr^2 + r^2 d\Omega^2_{d-1}   \ ,
\end{equation}
where 
\begin{equation}
 N(r)=1-\frac{2n(r)}{r^{d-2}}-\frac{2\Lambda}{(d-1)d} r^2  \ 
\end{equation}
and $d\Omega^2_{d-1}$ is the line element of a $(d-1)$-dimensional unit sphere. Note that the gravitational
constant is chosen such that the $d=2$ case has no Newtonian limit \cite{d2}.
As such the metric function $n(r)$ is well-behaved also in the $d=2$ case. Note that for 
weak and static gravitational fields $g_{tt}\sim -(1+2\psi(r))$,
where $\psi(r)$ is the Newtonian potential. Then the behaviour of $\psi(r)$ in $d=2$ which is
$\psi(r)\sim -\ln(r)$ would imply the divergence of $n(r)$ if that limit would exist within our 
Ansatz.

For the complex scalar field, we use a stationary Ansatz that contains a periodic dependence of the time-coordinate $t$:
\begin{equation}
\label{ansatz1}
\Phi(t,r)=e^{i\omega t} \phi(r) \ ,
\end{equation}
where $\omega$ is a constant and denotes the frequency.

In order to be able to use dimensionless quantities we introduce the following rescalings
\begin{equation}
\label{rescale}
 r\rightarrow \frac{r}{m} \ \ , \ \ \omega \rightarrow m\omega \ \ , \ \  \ell \rightarrow \ell/m \ \ ,
 \  \phi\rightarrow \eta_{\rm susy} \phi \ \ , \ \ n\rightarrow n/m^{d-2}  
\end{equation}
and find that the equations depend only on the dimensionless coupling constants
\begin{equation}
 \kappa=8\pi G_{d+1}\eta_{\rm susy}^2 = 8\pi \frac{\eta_{\rm susy}^2}{M_{\rm pl,d+1}^{d-1}}  \ ,
\end{equation}
where $M_{\rm pl,d+1}$ is the $(d+1)$-dimensional Planck mass. 
Note that with these rescalings the scalar boson mass $m_{\rm B}\equiv m$ 
becomes equal to unity. In these rescaled variables and coupling constants the coupled system of 
non-linear ordinary differential equations reads
\begin{equation}
 n'=\kappa\frac{r^{d-1}}{2}\left(N\phi'^2 + U(\phi) + \frac{\omega^2\phi^2}{A^2 N}\right)  \ ,
\end{equation}
\begin{equation}
 A'=\kappa r\left(A\phi'^2 + \frac{\omega^2\phi^2}{A N^2}\right)  \ ,
\end{equation}
\begin{equation}
\label{phi_eq}
 \left(r^{d-1} A N \phi'\right)'= r^{d-1} A 
\left(\frac{1}{2} \frac{\partial U}{\partial \phi} -\frac{\omega^2\phi}{N A^2}\right)  \ .
\end{equation}
These equations have to be solved numerically subject to appropriate boundary conditions.
We want to construct globally regular solutions with finite energy. At the origin we hence require
\begin{equation}
\label{bc1}
\phi'(0)=0 \ \ , \ \ \ n(0)=0 \ ,
\end{equation}
while we choose
$A(\infty)=1$ (any other choice would just result in a rescaling of the time coordinate). 
Moreover, while the scalar field function falls of exponentially for $\Lambda=0$  with
\begin{equation}
\label{bc2a}
 \phi(r>>1)\sim \frac{1}{r^{\frac{d-1}{2}}} \exp\left(-\sqrt{1-\omega^2}r\right) + ...
\end{equation}
it falls of power-law for
$\Lambda < 0$ with
\begin{equation}
\label{delta}
 \phi(r>>1)=\frac{\phi_{\Delta}} {r^{\Delta}} \ \ , \ \ 
\Delta=\frac{d}{2}+\sqrt{\frac{d^2}{4}+\ell^2}  \ .
\end{equation}
When solving the equations numerically, we will choose as fourth boundary condition
$\phi(\infty)=0$ for $\Lambda=0$, while for $\Lambda < 0$ we will choose the fall-off given in (\ref{delta}).
$\phi_{\Delta}$ is then a constant that has to be determined numerically and
which can be interpreted via the AdS/CFT correspondence as the value of the condensate of glueballs in the dual theory living
on the $d$-dimensional boundary of global AdS.

The explicit expression for the Noether charge reads
\begin{equation}
\label{charge_ex}
 Q=\frac{2 \pi^{d/2}}{\Gamma(d/2)} \int\limits_0^{\infty} {\rm d} r \ r^{d-1} \frac{\omega\phi^2}{AN} \ .
\end{equation}
For $\kappa\neq 0$, we can determine the mass $M$ from the behaviour of the metric function $n(r)$ at infinity. This reads \cite{radu}
\begin{equation}
 n(r \gg 1)= M + n_1 r^{2\Delta+d} + .... \ , 
\end{equation}
where $n_1$ is a constant that depends on $\ell$. 

For $\kappa=0$ we have $A\equiv 1$ and $n\equiv 0$. Then the mass 
$M$ corresponds to the integral of the energy density $T^0_0$ 
and reads
\begin{equation}
\label{mass}
 M=\frac{2 \pi^{d/2}}{\Gamma(d/2)} \int\limits_0^{\infty} {\rm d} r \ r^{d-1} \  \left(N \phi'^2 + 
\frac{\omega^2 \phi^2}{N} + U(\phi)\right)  \ .
\end{equation}
Note that this expression is perfectly finite in AdS. While $N$ contains a term $\propto r^2$ the fall-off of
the scalar function $\phi$ guarantees that $M$ has a finite value. Using the expression
for the charge $Q$ (\ref{charge_ex}), we can give a relation between the charge and the mass
\begin{equation}
 M=\omega Q + \frac{2 \pi^{d/2}}{\Gamma(d/2)} \int\limits_0^{\infty} {\rm d} r 
\ r^{d-1} \  \left(N \phi'^2 + U(\phi)\right)  \ .
\end{equation}
For $\Lambda=0$ it was found \cite{kusenko,ct} that in the ``thin-wall'' approximation which corresponds
to $\omega \sim 0$ the mass and charge are related as follows
\begin{equation}
 \frac{M}{\omega Q} \sim \frac{d+1}{d} \ .
\end{equation}
This has been used to draw conclusions on the stability of the $Q$-balls.
On the other hand, when $\omega$ becomes comparable to the scalar boson mass $m_{\rm B}\equiv m$ which 
corresponds to $\omega$ being close to its maximal possible value the $Q$-balls become very spread out, they
possess hence a small charge $Q$ 
and we can use the so-called ``thick-wall'' approximation \cite{kusenko,ct,multamaki_vilja}. This has been done for $Q$-balls
in Minkowski space-time with a self-interaction potential of the form $U(\phi) \sim m^2\phi^2 - a \phi^p$
for an arbitrary power $p$, arbitrary constant $a$ and in arbitrary spatial dimensions $d$. Since in the
thick wall approximation $\phi$ is very small we can approximate our exponential potential by this potential 
for $p=4$. The main result of \cite{multamaki_vilja} is that the following condition has to
be fulfilled in order to have stable $Q$-balls in the thick-wall limit: $4-2d > 0$. 
We see immediately that for all $d \geq 2 2$ this is not fulfilled and we would expect that the thick-wall $Q$-balls are 
unstable.
This was also found in \cite{ct} in $d=3$ and is confirmed by our numerical findings for general 
$d \geq 2$ (see Numerical results section below). Since a negative cosmological constant acts as
an attractive force, we would expect that $Q$-balls can become stable for sufficiently negative $\Lambda$ in
the thick-wall limit. Analytical results are very difficult to find here and are beyond the scope of this
paper. Our numerical results show that $Q$-balls in AdS can become stable for sufficiently large $Q$, i.e.
in the ``thick-wall limit'' (see below).

\section{Numerical results}
The solutions to the coupled system of nonlinear differential equations 
are only known numerically. We have solved these equations
using the ODE solver COLSYS \cite{colsys}. 
The solutions have relative errors on the order of $10^{-6}-10^{-10}$.
\subsection{$Q$-balls}
We have first studied the case $\kappa=0$. This corresponds to $Q$-balls in a Minkowski $(\Lambda=0$) 
or AdS background ($\Lambda < 0$), respectively. In this case, the Einstein
equations decouple from the system and $n(r)\equiv 0$, while $A(r)\equiv 1$.

It is known that $Q$-balls in $d=3$ exist on a limited interval
of the frequency $\omega\in [\omega_{\rm min}:\omega_{\rm max}]$. For $\Lambda=0$ the mass and charge diverge at
both boundaries \cite{kk1,kk2}, while for $\Lambda\neq 0$ this is still true at 
$\omega_{\min}$, but now
the two quantities tend to zero at $\omega_{\rm max}$ \cite{hartmann_riedel}. This is related to the fact that
in the limit $\omega\rightarrow \omega_{\rm max}$ the scalar field function
tends to zero everywhere.
In Minkowski space-time this still leads to an infinite value of 
the integral since the space-time is infinite,
however AdS space-time acts as a confining box and as such the integral becomes zero. 

\subsubsection{$\Lambda=0$}
$Q$-balls with an exponential interaction potential of the form (\ref{potential}) have been
studied in \cite{ct,hartmann_riedel} in $(3+1)$ dimensions. Here, we extend these results to $d\neq 3$
and show that the behaviour of the mass $M$ and the charge $Q$ depend crucially on the space dimension $d$.
  
Our results for the mass $M$ and charge $Q$ in dependence on $\omega$ are given 
in Fig.\ref{mq_om0} for $d=2,3,4,5,6$. We observe that the mass and charge diverge at $\omega=\omega_{\rm max}$ 
for $d\geq 3$, while for $d=2$ they tend to finite values. 
Moreover, since the potential and effective potential $U_{\rm eff}:=\omega^2 \phi^2 - U(\phi)$   
do not depend on $d$ the arguments employed in \cite{vw} can also be used here, such that
$\omega_{\rm max}\equiv 1$ and $\omega_{\rm min}\equiv 0$ do not depend on $d$. This is clearly
see in Fig.\ref{mq_om0}.

\begin{figure}[h]
\begin{center}

\subfigure[][$M$ over $\omega$]{\label{m_om0}
\includegraphics[width=5.5cm,angle=270]{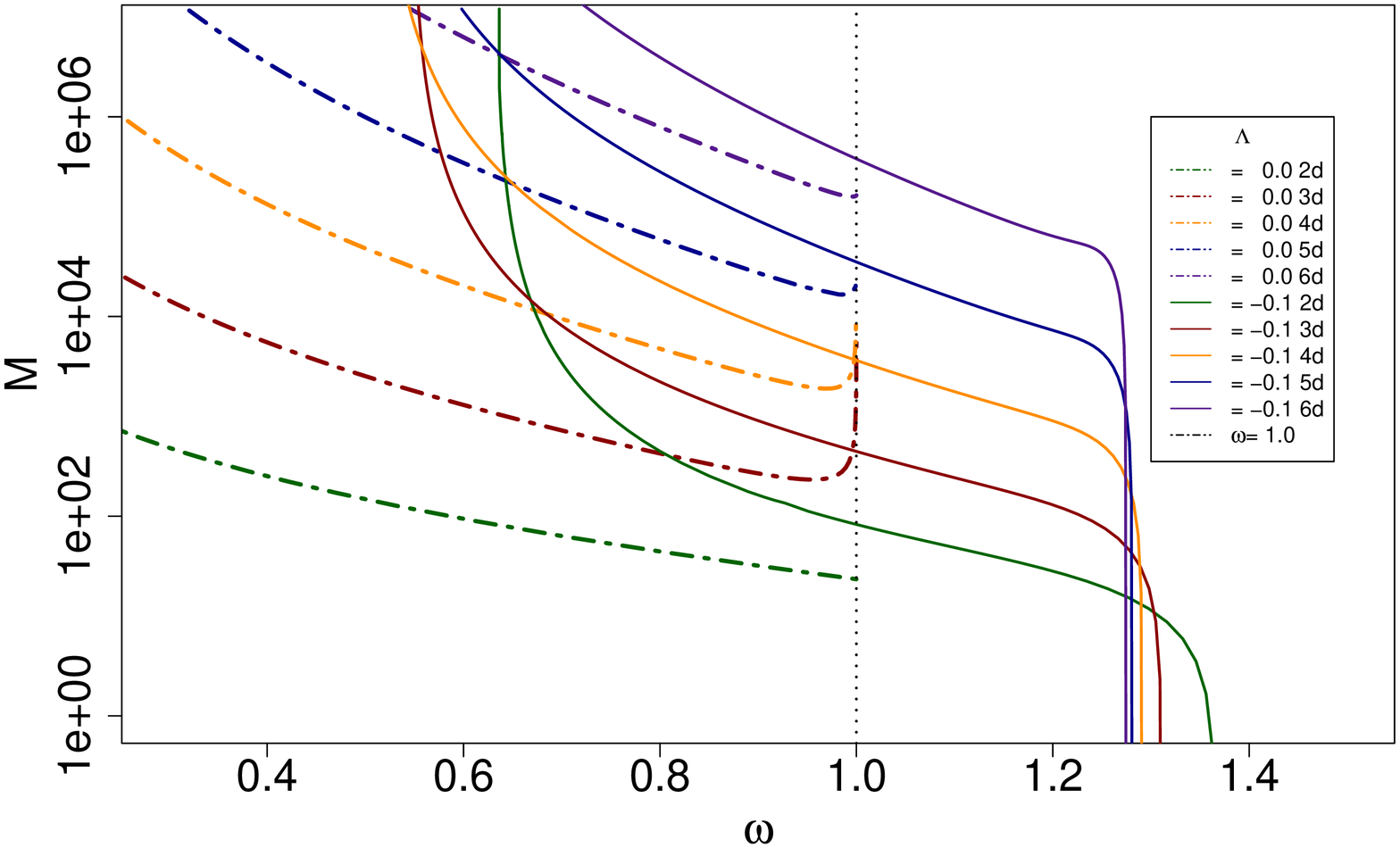}} 
\subfigure[][$Q$ over $\omega$]{\label{q_om0}
\includegraphics[width=5.5cm,angle=270]{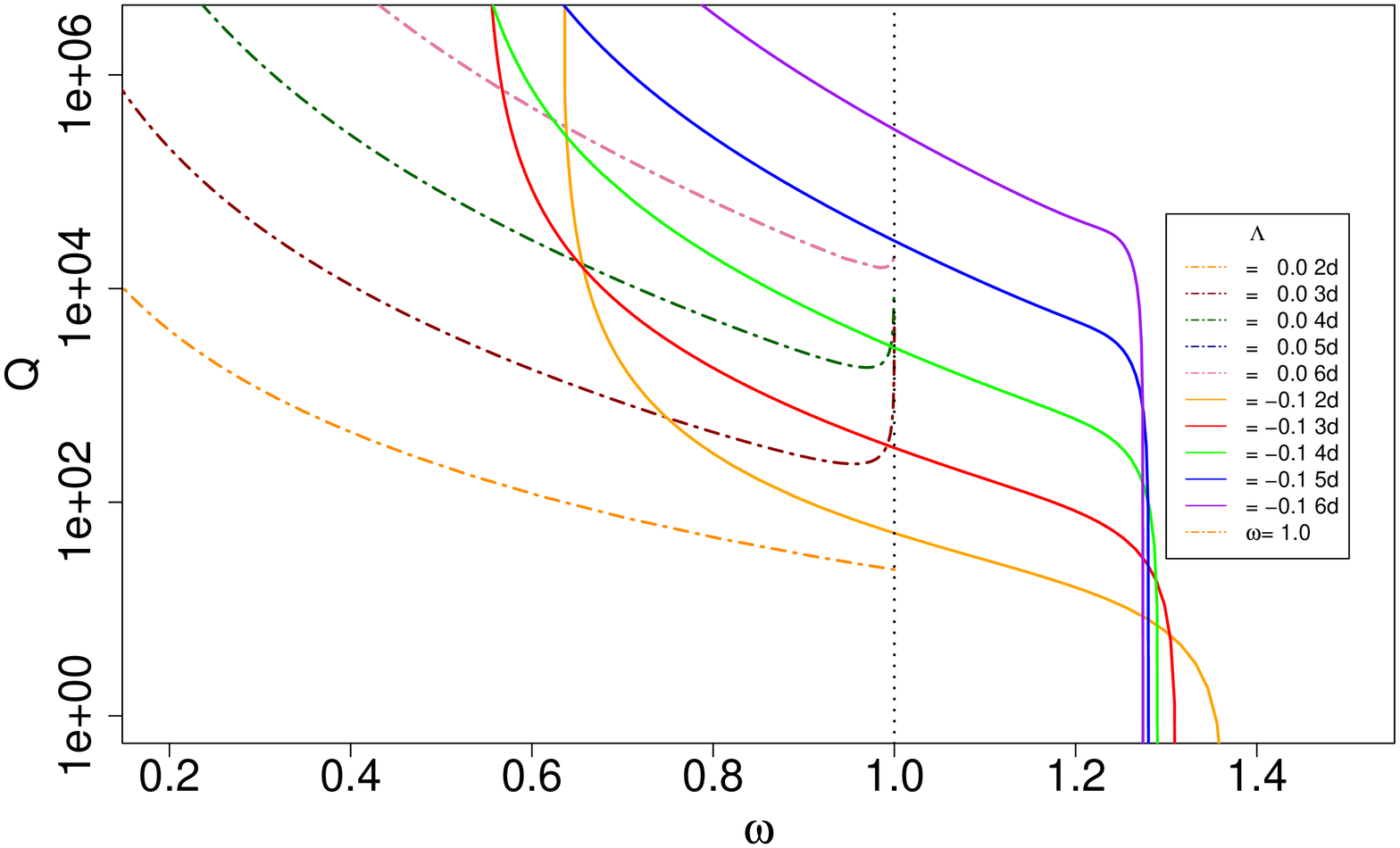}} 
\end{center}
\caption{\label{mq_om0} The value of the mass $M$ (left) and the charge $Q$ (right) of the $Q$-balls in dependence
on the frequency $\omega$ in Minkowski space-time ($\Lambda=0$) and AdS space-time $(\Lambda=-0.1$) 
for different values of $d$.}
\end{figure}

\begin{figure}[!htb]
\begin{center}
\includegraphics[width=8cm,angle=270]{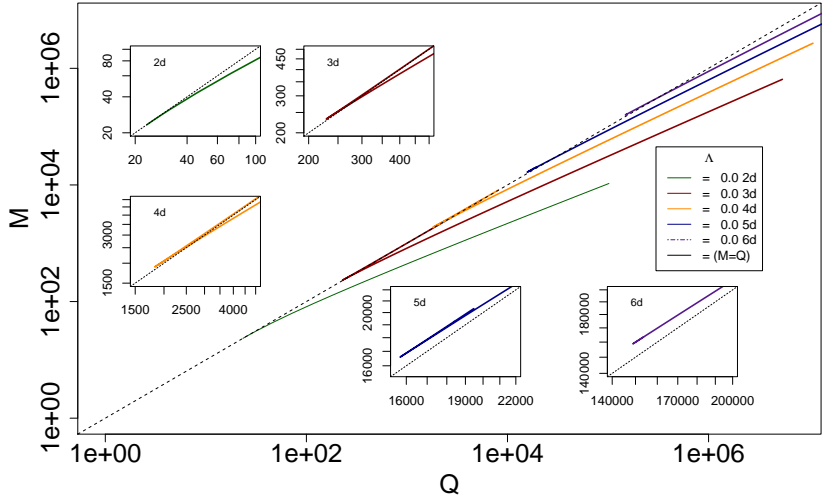}
\end{center}
\caption{\label{m_over_q_qballs} We show the value of the mass $M$ of the $Q$-balls  in dependence on their charge $Q$
for different values of $d$ in Minkowski space-time. The small subplots show the behaviour close to the minimal
value of $Q$.}
\end{figure}

In order to get an idea about the stability of these objects we can compare the mass $M$ with that
of $Q$ free scalar bosons of mass $m_{\rm B}\equiv m$. Due to our rescalings $m\equiv 1$ and
the mass of $Q$ free scalar boson is just equal to $Q$. Any solution with $M < Q$ would hence be stable
to decay into $Q$ free bosons. Our results for $d=2,3,4,5,6$ are shown in Fig.\ref{m_over_q_qballs}. 
In $d=3$ it was found \cite{kk1,kk2,hartmann_riedel} that there exist two solutions with different $M$ for a given charge $Q$,
one of which is stable to the decay into $Q$ free bosons, while the other is unstable. Here we observe that this
is also true for $d >3$: a stable branch exists up to a maximal value of the charge and then extends backwards to
form a second branch that for some critical value of the charge becomes unstable. We observe that the bigger $d$
the bigger is the value of the charge at which the solutions become unstable. 
On the other hand, for $d=2$ we find that the solutions are always stable with respect to the decay into
$Q$ free bosons. The fact that $Q$-balls are stable for small values of $\omega$ was pointed out already
in \cite{ct}. In this so-called ``thin-wall limit'' the $Q$-balls fulfill the relation
$M\simeq ((d+1)/d) \omega Q$. Since $\omega$ is small $M < Q$ in this limit
and the $Q$-balls are stable with respect to a decay to $Q$ 
free bosons. On the other hand for the ``thick-wall limit'' it is more difficult to
make analytical statements, but the results in \cite{ct} again agree with our numerical findings. This is
shown in Fig.\ref{ratio_m_q_over_omega}, where we show $M/Q$ as function of $\omega$. For $\omega$ close
to unity the $Q$-balls are in the ``thick-wall limit''. Clearly, the $Q$-balls are unstable in this case
since $Q < M$.

\begin{figure}[!htb]
\begin{center}
\includegraphics[width=8cm,angle=270]{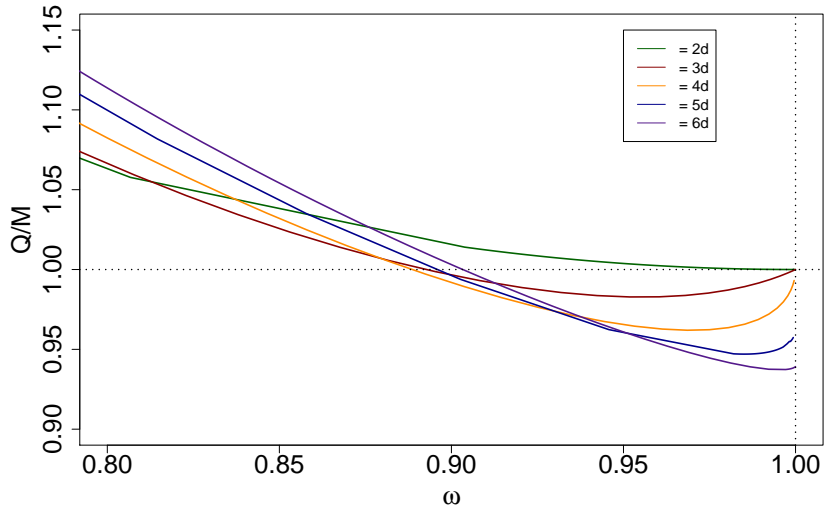}
\end{center}
\caption{\label{ratio_m_q_over_omega} We show the value of $Q/M$ in dependence on $\omega$ for 
$Q$-balls close to the ``thick-wall limit''. $Q$-balls are stable to decay into $Q$ scalar bosons of mass $m$
for $Q/M > 1$. Clearly, the $Q$-balls are unstable in the ``thick-wall limit'' for $d\geq 3$, while $M\rightarrow Q$
for $d=2$ in this limit.}
\end{figure}

\subsubsection{$\Lambda\neq 0$}
This case corresponds to $Q$-balls in a fixed AdS background and has been studied for the exponential potential
and $d=3$ in \cite{hartmann_riedel}. Our results are shown in Fig.\ref{m_over_q_qballs} for $\Lambda=-0.1$ and $d=2,3,4,5,6$.
Similar to $d=3$ the value of the mass $M$ and charge $Q$ tend to infinity at $\omega_{\rm min}=0$ independent
of $d$. On the other hand, the mass and charge tend to zero at $\omega_{\rm max}$. 
Moreover, we find that $\omega_{\rm max}$ decreases with increasing $d$ having the largest value for $d=2$. 
For $\Lambda \neq 0$ and $\kappa=0$ it is known that in the $d=3$ case and a particular choice of potential
exact solutions to the scalar field equation exist \cite{radu_subagyo}. We show in Appendix 1 that 
this generalizes to $d$ dimensions.
While the potential necessary to obtain this result is not of the form chosen in this paper, however,
our numerical results for $\omega_{\rm max}$ agree quite well with the analytic expression given by
$\omega_{\rm max}=\Delta/\ell$. This is related to the fact that for $\omega \rightarrow \omega_{\rm max}$ the
function $\phi(r)$ tends to zero everywhere. Hence all higher order terms in the potential become
negligible and the Ansatz made in \cite{radu_subagyo} gives a good result. 

\begin{figure}[h]
\begin{center}

\subfigure[][$\omega_{\rm max}$ over $\Lambda$]{\label{om_lam}
\includegraphics[width=5.5cm,angle=270]{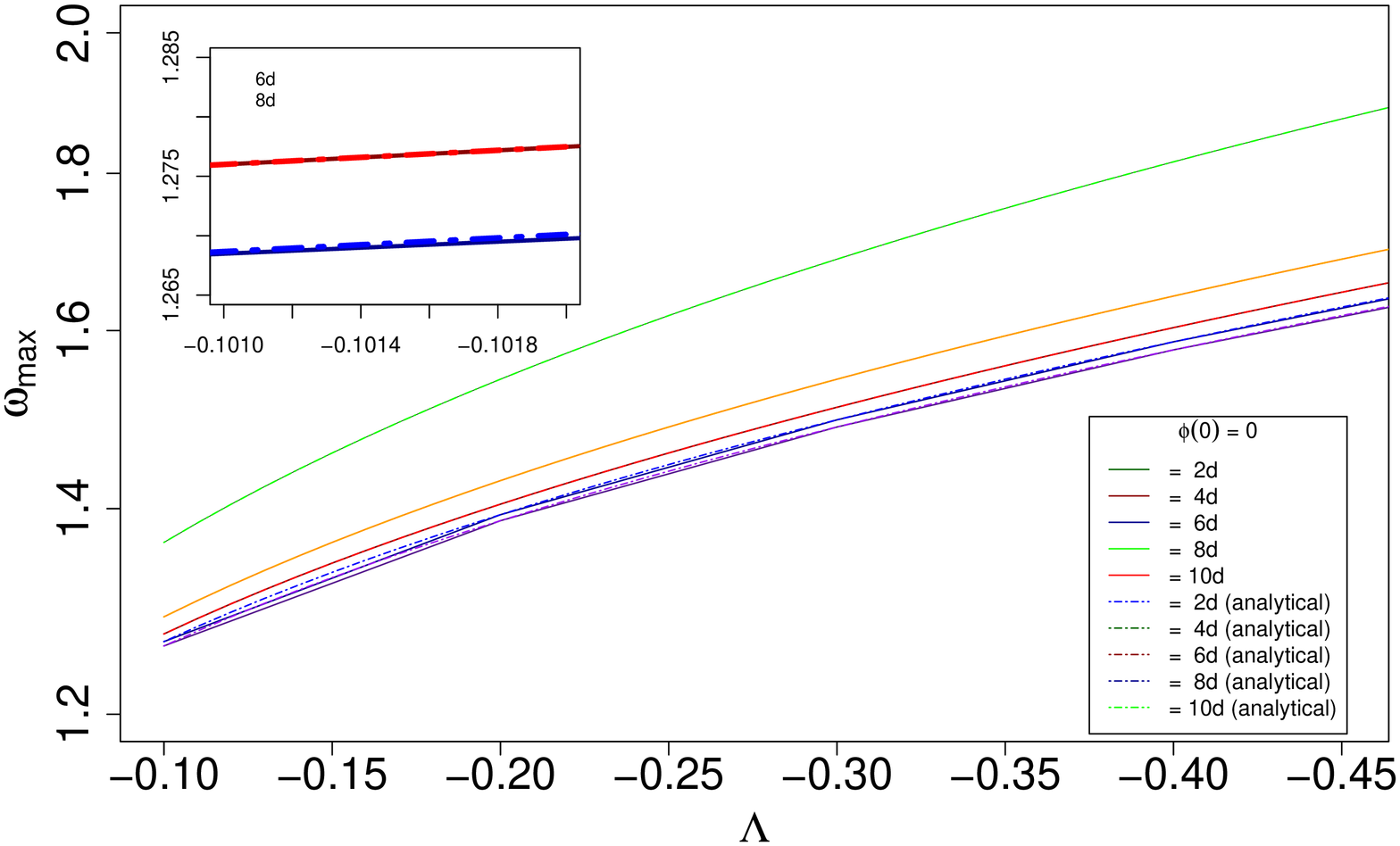}} 
\subfigure[][$\omega_{\rm max}$ over $d$]{\label{om_d}
\includegraphics[width=5.5cm,angle=270]{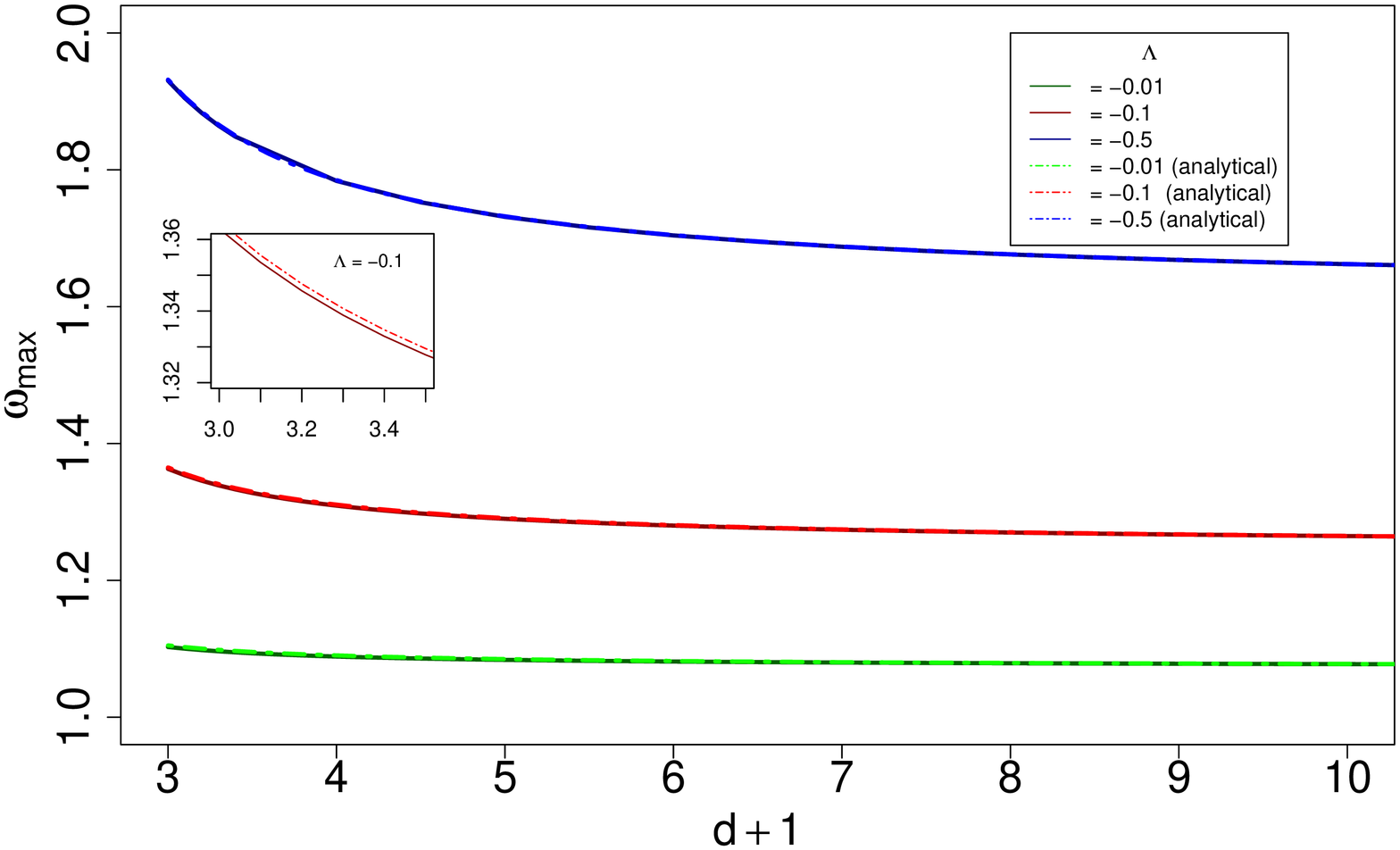}} 
\end{center}
\caption{\label{omega_max} The value of $\omega_{\rm max}$ in dependence on $\Lambda$ (left) and in dependence
on $d$ (right). Though we plot $d$ here as a continuous parameter, we should only read of the value for
$d\in \mathbb{N}$. We also give the value of $\Delta/\ell$ and find that it gives a good approximation
to our numerical data. }
\end{figure}

Our numerical results for $\omega_{\rm max}$ in dependence
on $\Lambda$ and $d$ are shown in Fig.\ref{omega_max} together with the value of $\Delta/\ell$.
As is apparent from this figure the analytical result agrees quite well with our numerical
values. Moreover, we observe as expected from the analytical result that $\omega_{\rm max}$ increases with decreasing $\Lambda$ 
and decreases with increasing $d$. 

In Fig.\ref{m_over_q_qballs_lam} we show the value of the mass $M$ in dependence on $Q$ for $d=2,3,4,5,6$ 
and $\Lambda=-0.1$, while in Fig.\ref{ratio_m_q_over_omega} we present 
$Q/M$ in dependence on $Q$ for the ``thick-wall limit''.  
Very similar to $d=3$ the $Q$-balls have mass $M$ larger than $Q$ and are hence unstable. This is clearly
see in Fig.\ref{ratio_m_q_over_omega}, while for
sufficiently large $Q$ they become stable with respect to this decay. The value of $Q=Q_{\rm crit}$ at which 
this transition happens depends on $\Lambda$ and it was found that $Q_{\rm crit}$ increases with decreasing 
$\Lambda$ \cite{hartmann_riedel}.
We find that the number of spatial dimensions $d$ also has an influence on the value of $Q_{\rm crit}$. 
We find that $Q_{\rm crit}$ increases with increasing $d$.

\begin{figure}[!htb]
\begin{center}
\includegraphics[width=8cm,angle=270]{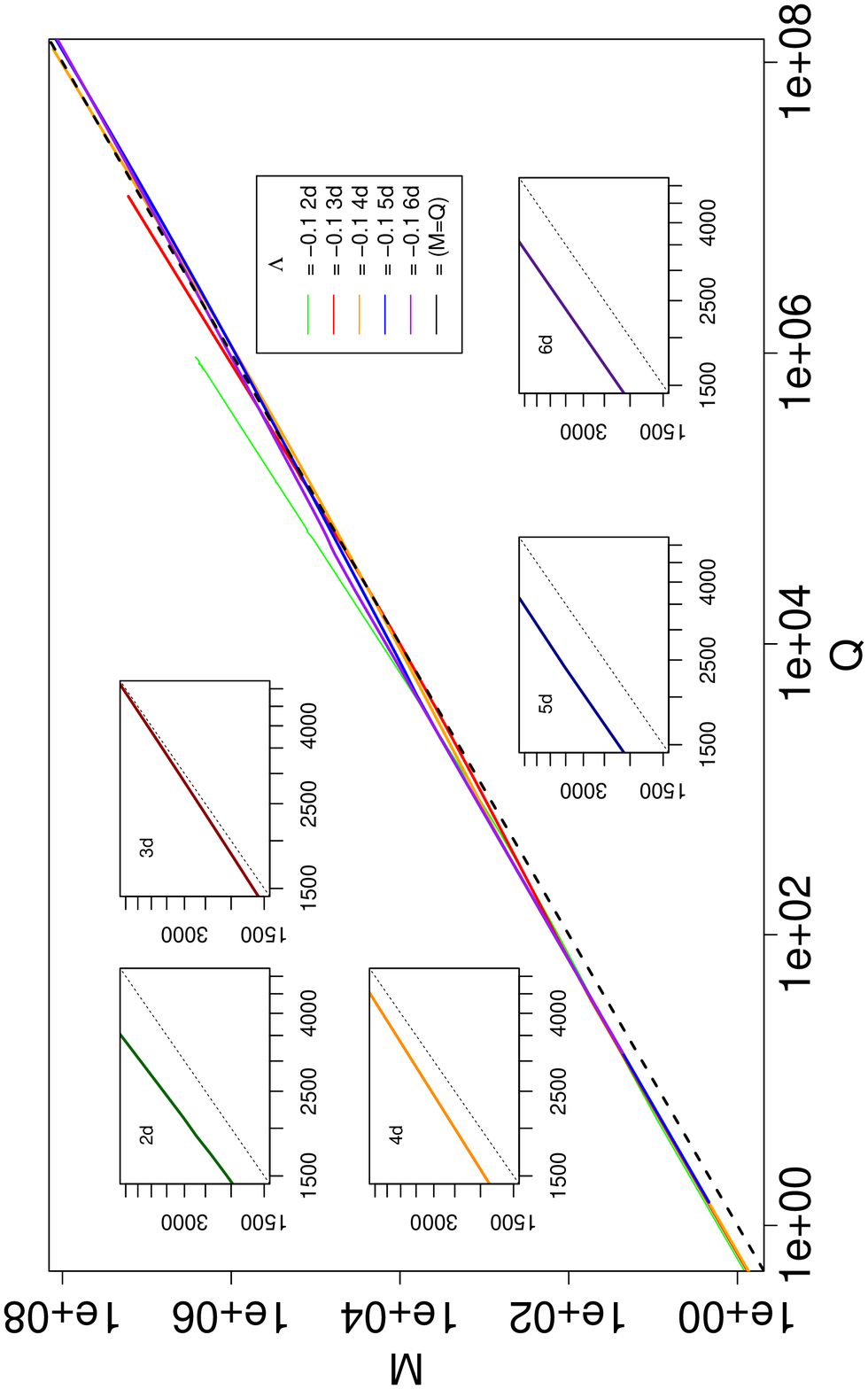}
\end{center}
\caption{\label{m_over_q_qballs_lam} We show the value of the mass $M$ of the $Q$-balls  in dependence on their charge $Q$
for different values of $d$ in AdS space-time with $\Lambda=-0.1$. }
\end{figure}

\begin{figure}[!htb]
\begin{center}
\includegraphics[width=8cm,angle=270]{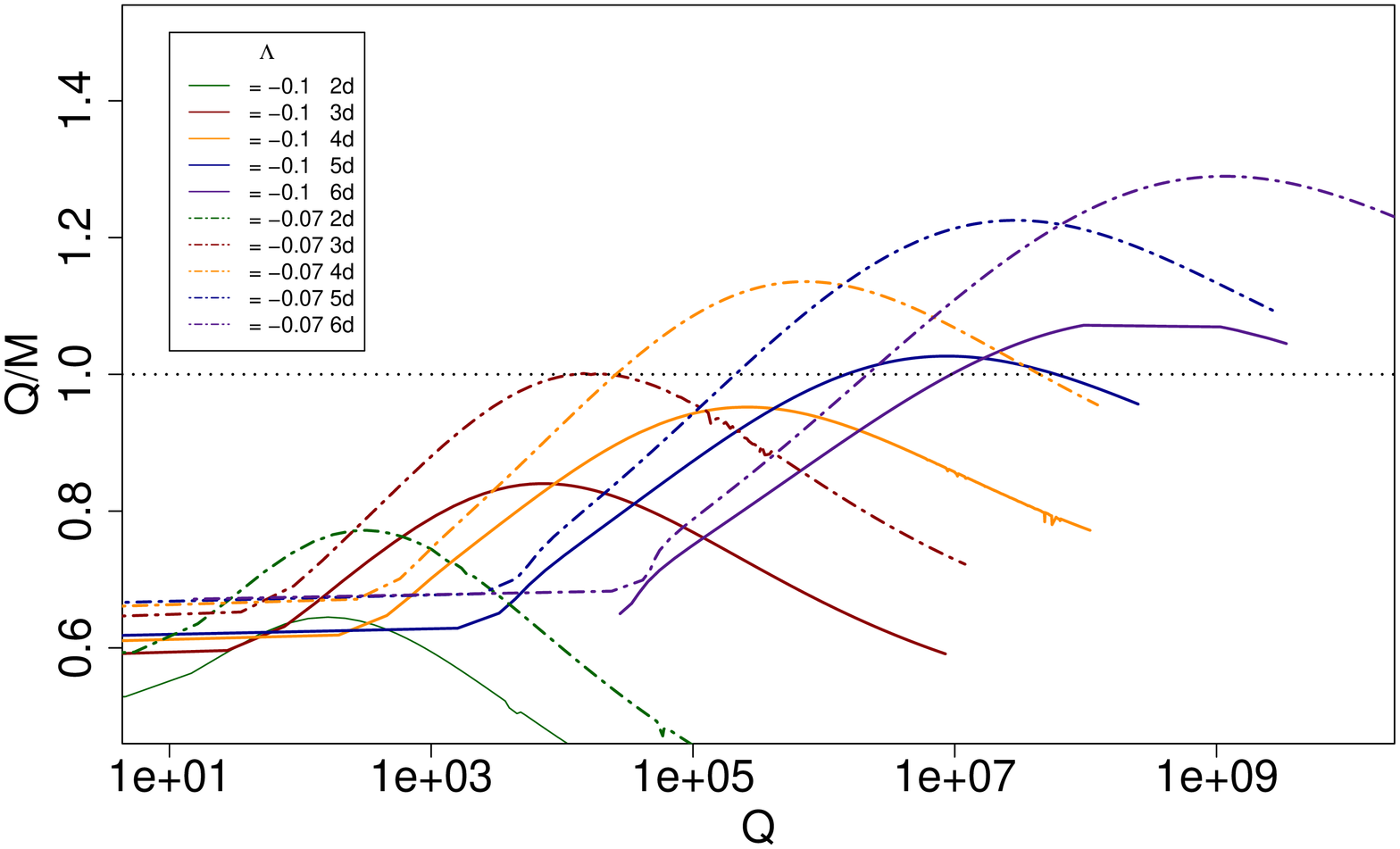}
\end{center}
\caption{\label{ratio_m_q_over_q} We show the value of $Q/M$ in dependence on $Q$ for 
$Q$-balls close to the ``thick-wall limit''. $Q$-balls are stable to decay into $Q$ scalar bosons of mass $m$
for $Q/M > 1$. Clearly, the $Q$-balls can become stable in the ``thick-wall limit'' for 
sufficiently large $d$ and/or $\vert\Lambda\vert$. }
\end{figure}

It is also known that radially excited $Q$-ball solutions exist which possess a number $k\in \mathbb{N}$ of zeros in
the scalar field function. In Fig.\ref{m_om_ex1} we show the mass $M$ as function of $\omega$ and as function of $Q$, respectively
for $\Lambda=-0.1$ and $d=3,4$ and $k=0,1,2$. We observe that for fixed $d$ the value of $\omega_{\rm max}$ increases with the
increase of $k$. Moreover, the bigger $k$ the bigger is the difference between $\omega_{\rm max}$ for 
$d=3$ and $d=4$. The dependence of the mass $M$ on $Q$ shown in Fig.\ref{q_om_ex1} indicates that all
solutions with nodes are unstable to decay into $Q$ free bosons. This is not surprising since these
can be seen as excited $Q$-balls in AdS space-time.

As pointed out in \cite{horowitz} the field theory on the boundary of AdS describes
condensates of scalar glueballs. This was further investigated in \cite{hartmann_riedel}, where
$Q$-balls in $(3+1)$-dimensional asymptotically global AdS have been studied. In Fig.\ref{qball_expectation}
we show our results for different values of $\Lambda$ and $d$. Apparently, the expectation
value of the dual operator $<{\rm O}>^{1/\Delta}$, which corresponds to the value of the condensate
of scalar glueballs decreases for increasing $d$ when fixing $\phi(0)$. This is related to the fact that
the scalar field can spread into more dimensions when $d$ is increased and hence less condensate is collected.
Furthermore, the value of the condensate increases with decreasing $\Lambda$. This is connected to the fact
that the ``AdS box'' decreases in size for decreasing $\Lambda$ and as such the value of the condensate becomes
bigger.

\begin{figure}[h]
\begin{center}

\subfigure[][$M$ over $\omega$]{\label{m_om_ex1}
\includegraphics[width=5.5cm,angle=270]{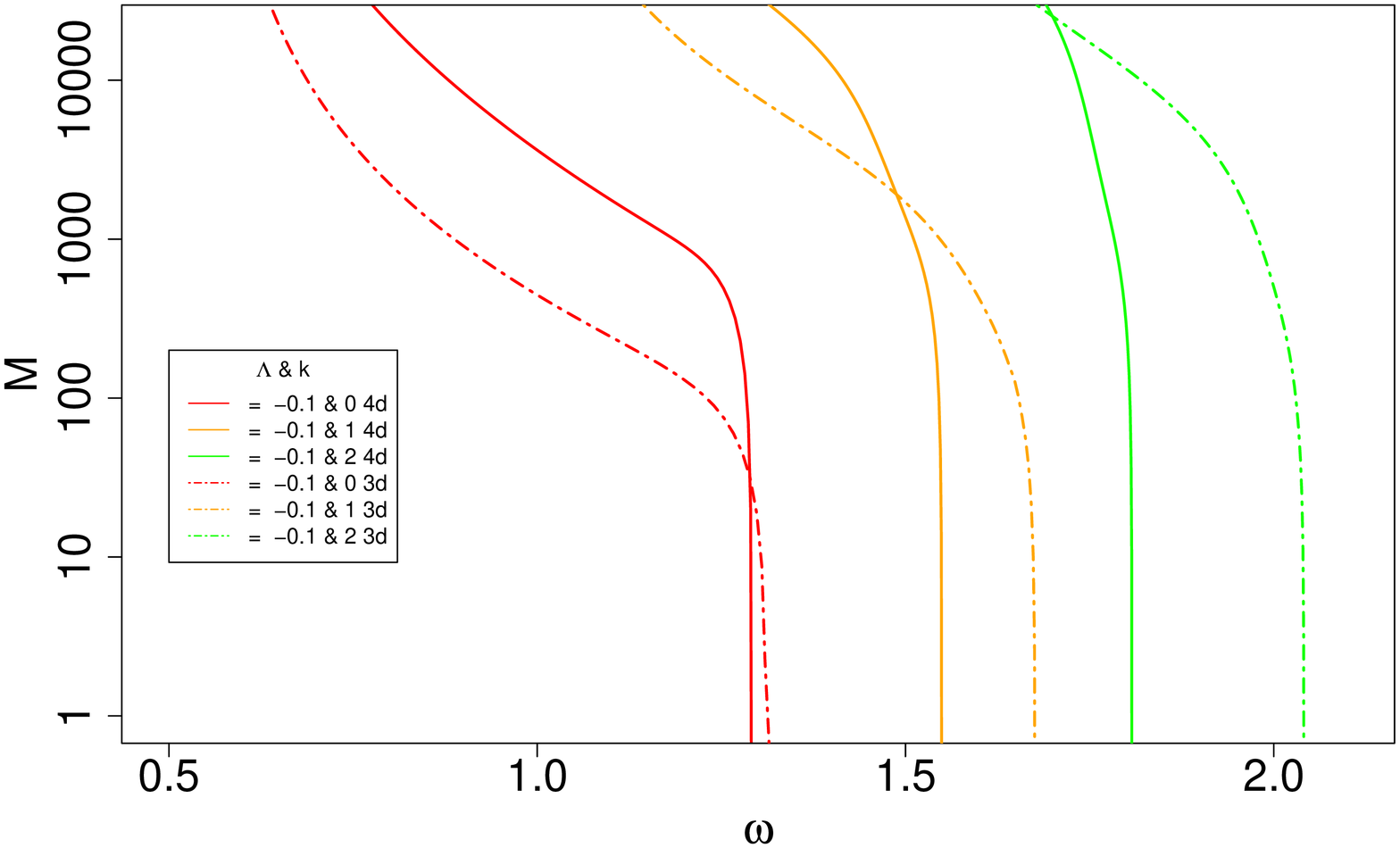}} 
\subfigure[][$M$ over $Q$]{\label{q_om_ex1}
\includegraphics[width=5.5cm,angle=270]{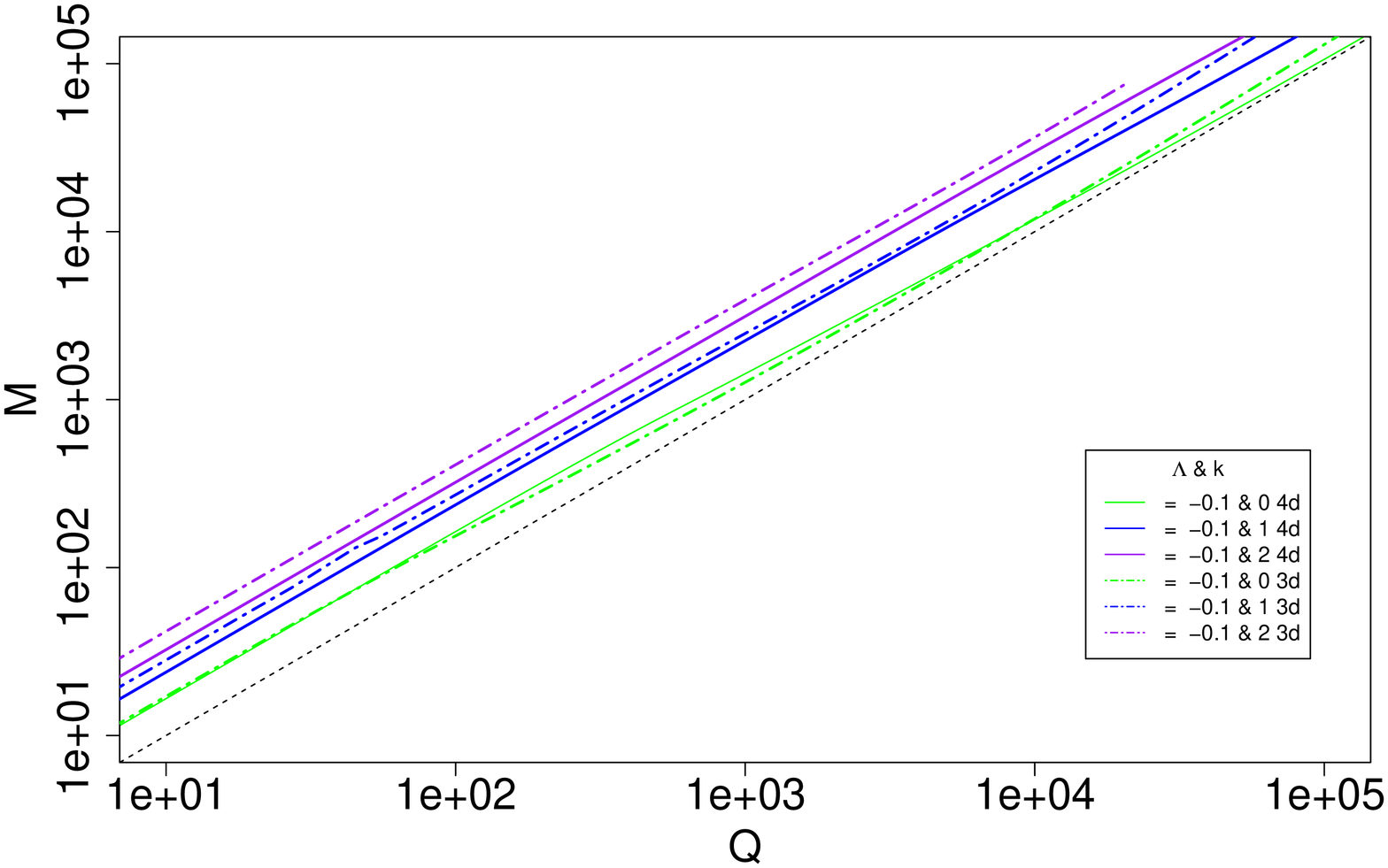}} 
\end{center}
\caption{\label{mq_om_ex1} The value of the mass $M$ of the $Q$-balls in dependence on $\omega$ (left) and in dependence 
on the charge $Q$ (right) in AdS space-time 
for different values of $d$ and number of nodes $k$ of the scalar field function. }
\end{figure}

\begin{figure}[!htb]
\begin{center}
\includegraphics[width=8cm,angle=270]{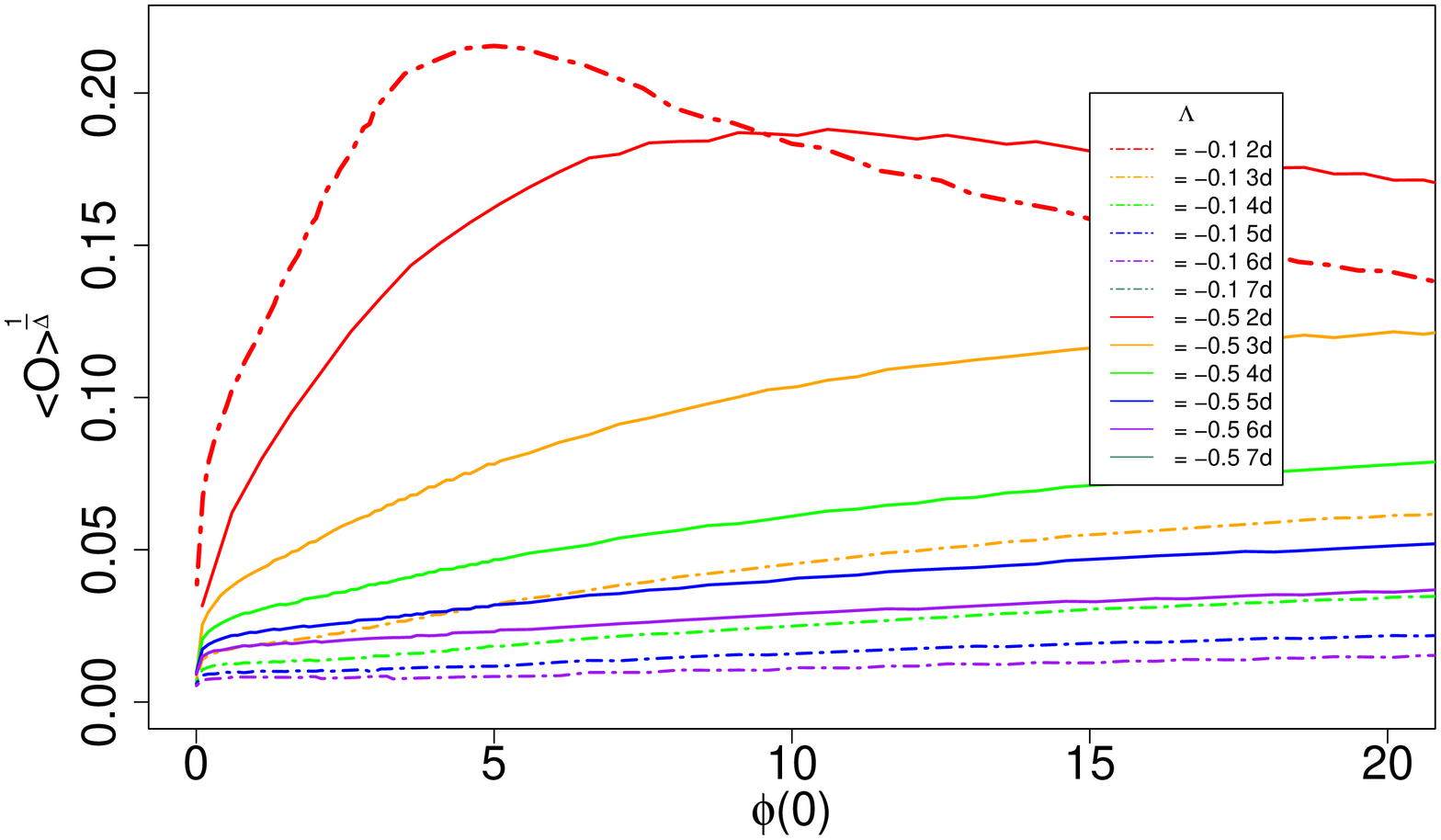}
\end{center}
\caption{\label{qball_expectation} We show the expectation value of the dual operator on the AdS boundary
$<{\rm O}>^{1/\Delta}$ corresponding to the value of the condensate of scalar glueballs 
in dependence on $\phi(0)$ for different values of $\Lambda$ and $d$.  }
\end{figure}

\subsection{Boson stars}
We now discuss the case $\kappa\neq 0$. This corresponds to boson stars in an asymptotically flat $(\Lambda=0$) 
or asymptotically AdS space-time ($\Lambda < 0$), respectively. 

\subsubsection{$\Lambda=0$}
As pointed out in \cite{radu}, boson stars in $(2+1)$-dimensional, asymptotically flat space-time do not
exist for massive scalar fields without self-interaction. In the Appendix 2 we show that this is different
in our case and that gravitating, asymptotically flat boson star solutions in $(2+1)$ do exist in our model with an
exponential self-interaction potential.

\begin{figure}[h]
\begin{center}
\includegraphics[width=8cm,angle=270]{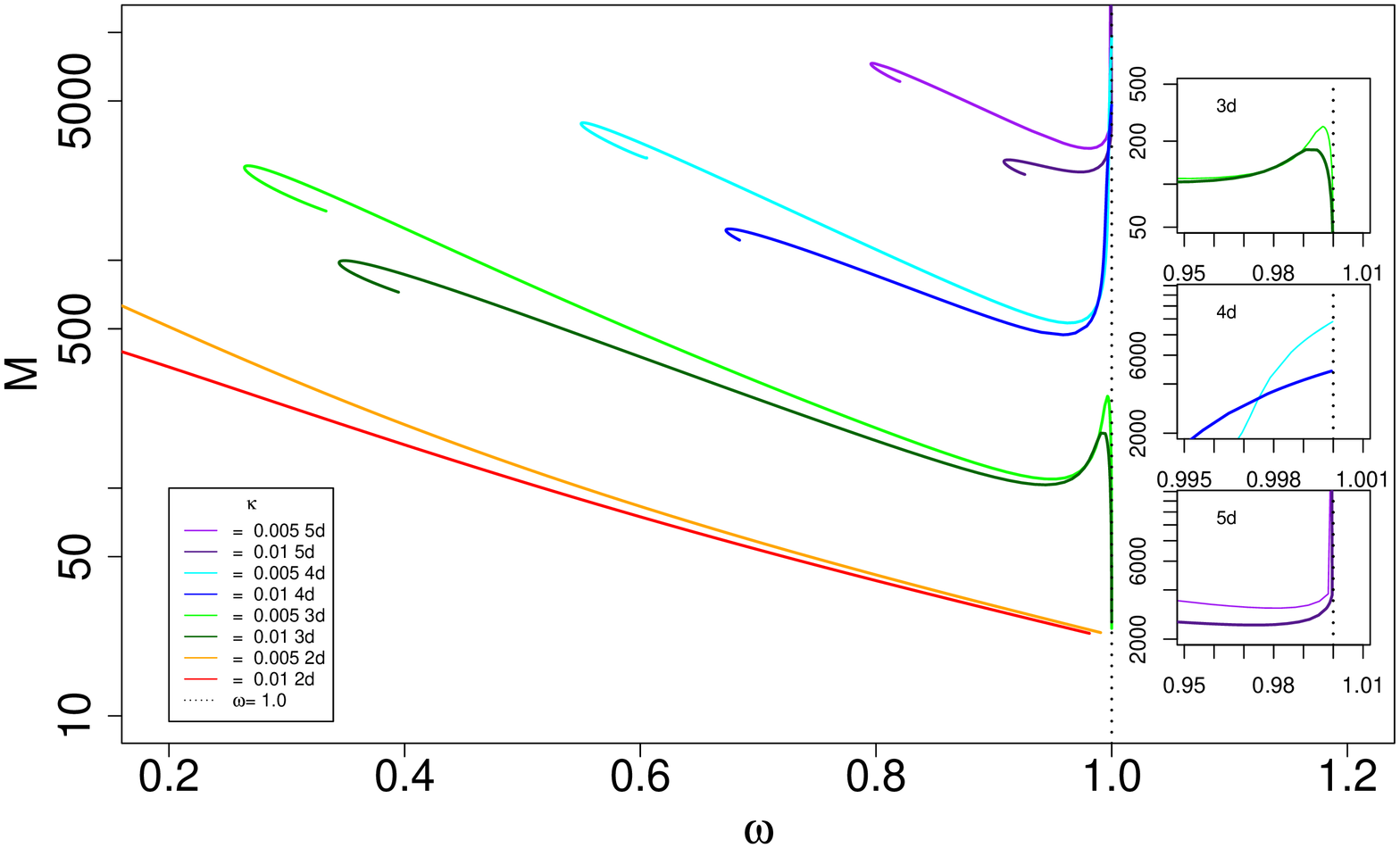}
\end{center}
\caption{\label{mq_om1} The value of the mass $M$ of the boson stars in dependence
on the frequency $\omega$ for $\Lambda=0$ and different values of $d$ and $\kappa$. The small
subfigures show the behaviour of $M$ at the approach of $\omega_{\rm max}$ for $d=3,4,5$ (from
top to bottom). Note that the curves for the charge $Q$ look qualitatively very similar, this is
why we don't give them here.}
\end{figure}

\begin{figure}[!htb]
\begin{center}
\includegraphics[width=8cm,angle=270]{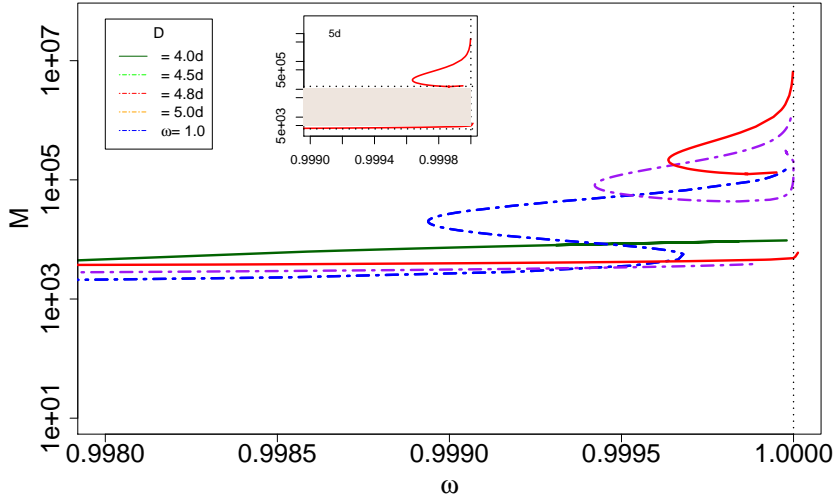}
\end{center}
\caption{\label{m_over_om_detail} We show the value of the mass $M$ of the boson stars in asymptotically
flat space-time ($\Lambda=0$) in dependence on the frequency $\omega$ close to $\omega_{\rm max}$. 
Note that only $d=4$ and $d=5$ are physical values, but that the dimension is a parameter in
our numerical programme that can also have non-integer values. Here we demonstrate how the mass evolves when
going from $d=4$ to $d=5$. The subplot further demonstrates that there exists a mass gap in $d=5$.}
\end{figure}

\begin{figure}[!htb]
\begin{center}
\includegraphics[width=8cm,angle=270]{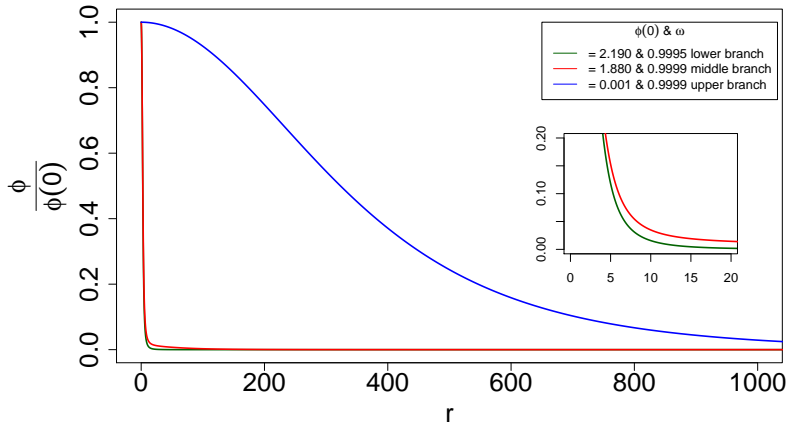}
\end{center}
\caption{\label{profiles} We show the profiles of the scalar field function $\phi(r)/\phi(0)$
for the case where three branches of solutions exist close to $\omega_{\rm max}$ in $d=5$. Here $\kappa=0.001$.}
\end{figure}

\begin{figure}[!htb]
\begin{center}
\includegraphics[width=8cm,angle=270]{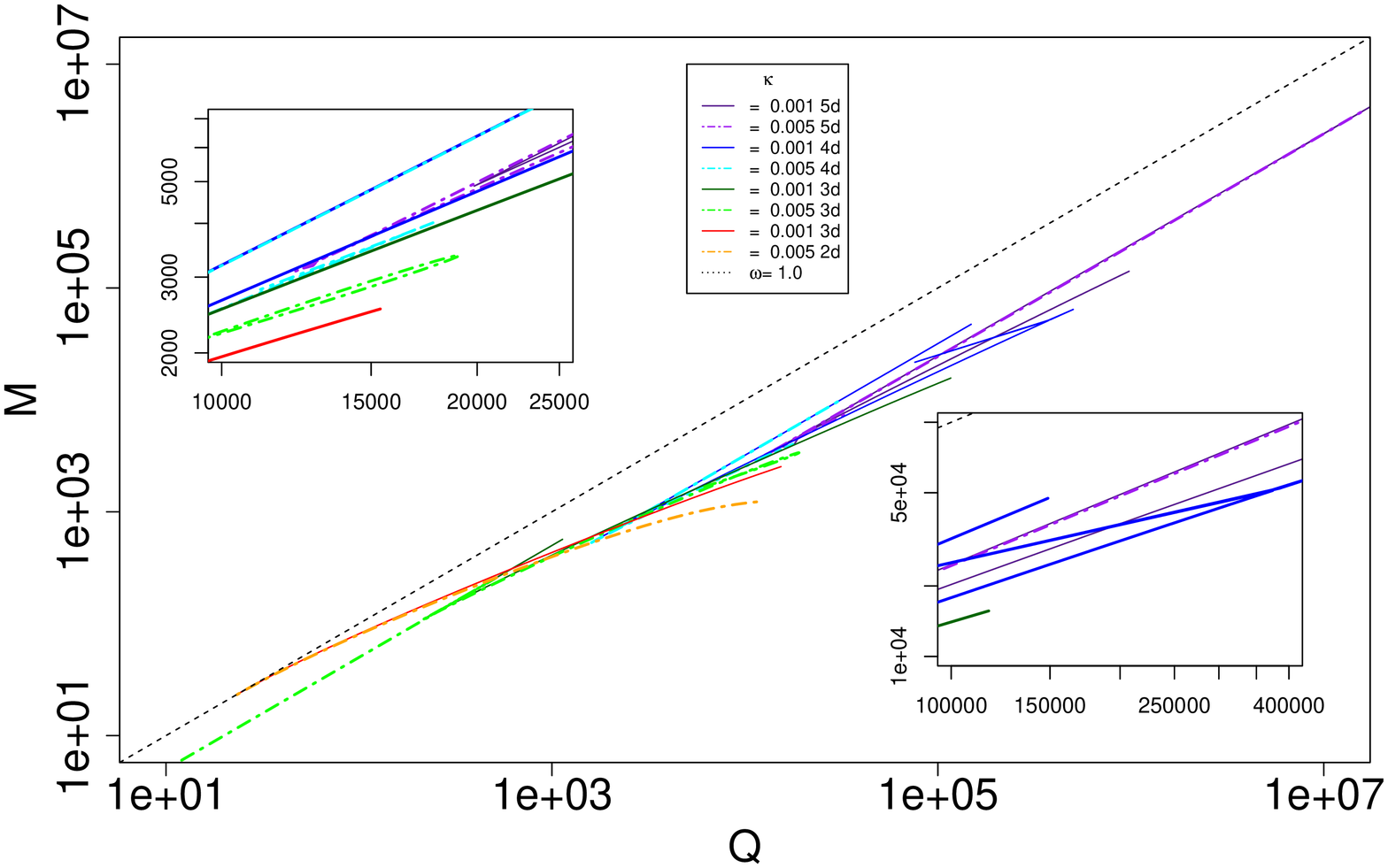}
\end{center}
\caption{\label{m_over_q} We show the value of the mass $M$ of the boson stars in asymptotically
flat space-time ($\Lambda=0$) in dependence on their charge $Q$
for different values of $\kappa$ and $d$. }
\end{figure}

We have also studied the dependence of the mass and charge on the frequency $\omega$. Our results are show
in Fig.\ref{mq_om1} for the mass. The curves look qualitatively similar for the charge $Q$, this is
why we don't show them here. We observe that the behaviour at $\omega_{\rm max}$ depends crucially on the number of
spatial dimensions $d$. For $d=3$ the mass and charge tend to zero, while for $d=4$ they tend to a finite
value. This has already been observed before and is confirmed with our type of potential. For $d=5$ we find
that now the mass and charge tend to infinity at the approach of $\omega_{\rm max}$. We have 
integrated up to values of the mass and charge of $10^7$ and believe that approaching $\omega_{\rm max}$ even closer,
these values would further increase. This can be understood
using the argument employed for $d=4$ in \cite{hartmann_kleihaus_kunz_list}. 
As noticed in this latter paper, the scalar field function and radial coordinate show a scaling behaviour
that is equal in $d=3$ and $d=4$. We find that this is also the case here and generalizes to $d > 4$ such that the behaviour is
$ \phi(r) \rightarrow \hat{\phi}(\hat{r})$, $n\rightarrow 0$, $A\rightarrow 1$ 
for $\omega\rightarrow \omega_{\rm max}$ with
\begin{equation}
 \hat{\phi}=\phi/\phi_0 \ \ , \ \ \hat{r}=\left(\phi_0 \kappa^{1/2}\right)^{1/2} r 
\end{equation}
with $\phi_0$ a constant that tends to zero in the limit $\omega\rightarrow \omega_{\rm max}$. Now this implies
e.g. for the charge $Q$ (the argument works similarly for the mass $M$):
\begin{equation}
\label{charge_scaling}
 Q=\frac{2 \pi^{d/2}}{\Gamma(d/2)} \omega \int\limits_0^{\infty} {\rm d} r \ \ r^{d-1} \ \phi^2 \ \ \longrightarrow \ \ 
Q=\frac{2 \pi^{d/2}}{\Gamma(d/2)} \omega_{\rm max} \phi_0^{2-d/2} \kappa^{-d/4} \int\limits_0^{\infty} 
{\rm d} \hat{r}  \ \hat{r}^{d-1} \ \hat{\phi}^2  \ .
\end{equation}
For $d\leq 3$ this tends obviously to zero, for $d=4$ this becomes constant,
and for $d\geq 5$ this tends to infinity, respectively for $\phi_0\rightarrow 0$.
In addition to this we observe that the approach to $\omega_{\rm max}$ is not smooth in $d=5$.
This is shown in Fig. \ref{m_over_om_detail}. For $d=3$ and $d=4$ the mass tends smoothly to zero and a finite
value, respectively. For $d=5$ we observe that the mass tends to a finite value on a lower branch of solutions,
but that close to $\omega_{\rm max}$ new branches of solutions exist. These are quite small
in extend and in fact are barely noticeable for $d > 5$. As such, a second branch of solutions extends backwards from
$\omega_{\rm max}$ down to a critical value of $\omega$ and then bends backwards to tend to infinity.
Our conclusion hence is that while these solutions can exist for arbitrarily large values of the mass (and charge)
there exists a mass gap in which solutions are not allowed. Furthermore, there is a small interval
of $\omega$ in which up to three solutions with different masses exist. To understand this pattern,
we plot the three solutions for $\omega$ close to $\omega_{\rm max}$ in Fig.\ref{profiles}. We observe that for the same value of
$\omega$ the three solutions are distinguished by the value of $\phi(0)$ with $\phi(0)$ decreasing from the first to the third branch.
Moreover, the solution spreads out more and more over $r$. On the first branch, the solution is still quite localized around the origin,
while it becomes very delocalised on the third branch. 

\begin{figure}[!htb]
\begin{center}
\includegraphics[width=8cm,angle=270]{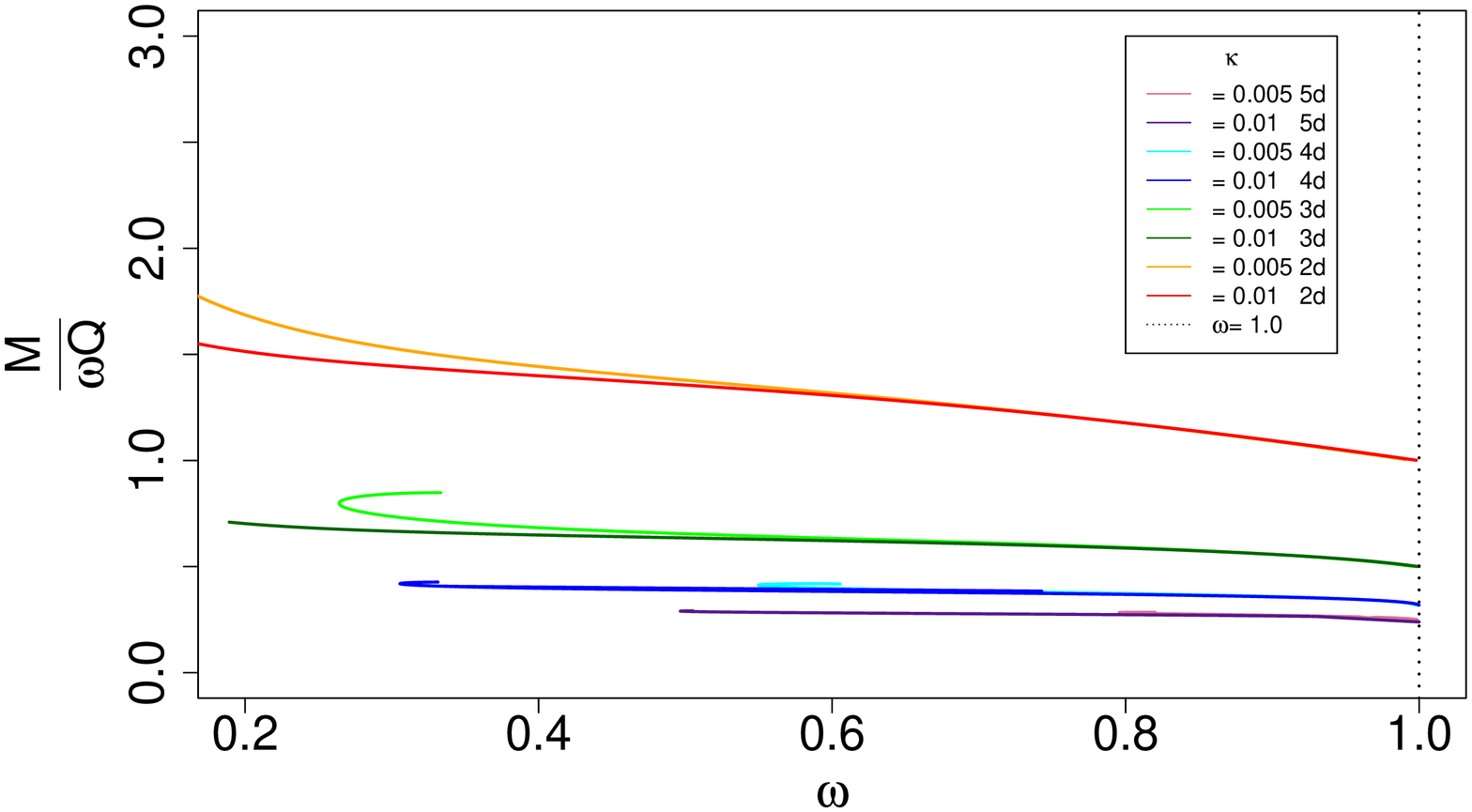}
\end{center}
\caption{\label{m_over_omegaq} We show the value of the mass $M/(\omega Q)$ 
of the boson stars in asymptotically
flat space-time ($\Lambda=0$) in dependence on the frequency $\omega$ for different values of $\kappa$ and $d$. }
\end{figure}

We further observe that for all $d$ the solutions exist down to a minimal value of the frequency $\omega_{\rm min}$.
At this value of $\omega$, a second branch of solutions exists that has lower mass and charge for a fixed value
of $\omega$. This second branch extends back to a critical value of $\omega$ and then forms another branch.
This spiraling has been observed before in $d=3$ and is apparently also present in $d\neq 3$. 

We can also read off the dependence of $\omega_{\rm min}$ and $\omega_{\rm max}$ on $d$ and $\kappa$ from
Fig.\ref{mq_om1}. As is apparent, $\omega_{\rm max}=1$ does neither depend on $\kappa$ nor on $d$. On the other
hand, the value of the minimal frequency $\omega_{\rm min}$ depends strongly on $\kappa$ and $d$.
For fixed $d$ it increases with increasing $\kappa$, i.e. the stronger the interaction between the gravitational
field and the scalar field the larger we have to choose the value of the frequency to find solutions.
This is true for  all $d$ that we have studied. We also notice that the decrease in $\omega$
is bigger in higher $d$ when increasing $\kappa$ by the same amount. For fixed $\kappa$ our results
indicate that $\omega_{\rm min}$ increases with increasing $d$.
Normally, we would expect gravity in higher $d$ 
to become weaker since it can leak into extra dimensions, however, here we are keeping $G_{\rm d+1}$, i.e. 
the strength of the gravitational interaction
constant.

We show the dependence of $M$ on $Q$ for different values of $\kappa$ and $d$ in Fig.\ref{m_over_q}.
As expected boson stars in asymptotically flat space-time are stable to decay into $Q$ free bosons
since these objects are gravitationally bound. This is also true for the case $d\geq 5$, where no bound on
the mass $M$ and charge $Q$ exist. As is clearly seen from Fig.\ref{m_over_q} we find that also
for $d=5$ the curve is always below $M=Q$. 

For $\kappa=0$ it was found that for small values of $\omega$ there is a relation between $M$ and $Q$ that depends
on $\omega$ and $d$ \cite{ct}. In Fig.\ref{m_over_omegaq} we plot $M/(\omega Q)$ 
as function of $\omega$ and find that only for $d=2$ the approximation of flat space-time $M/(\omega Q)\sim (d+1)/d$
is a good approximation for small $\omega$. For larger values of $d$ the solutions do not exist
for small $\omega$ and $M/(\omega Q)$ is always smaller than one.

\subsubsection{$\Lambda\neq 0$}
We have also studied boson stars in AdS space-time. Our results for the mass $M$ and charge $Q$ in dependence
on the frequency $\omega$ are shown in Fig.\ref{mq_om_lam} for $\Lambda=-0.1$.
 
\begin{figure}[h]
\begin{center}

\subfigure[][$M$ over $\omega$]{\label{m_om0}
\includegraphics[width=5.5cm,angle=270]{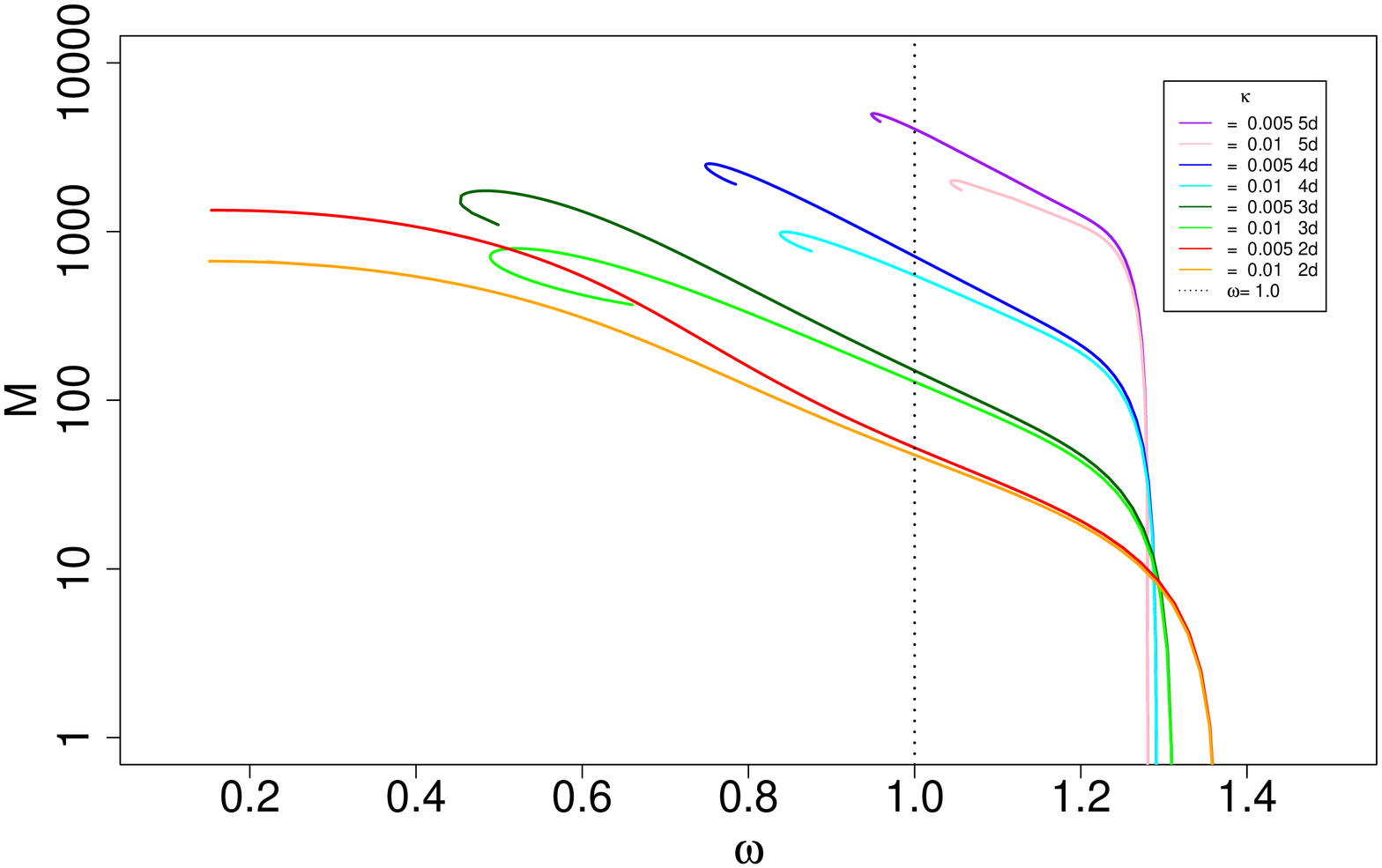}} 
\subfigure[][$Q$ over $\omega$]{\label{q_om0}
\includegraphics[width=5.5cm,angle=270]{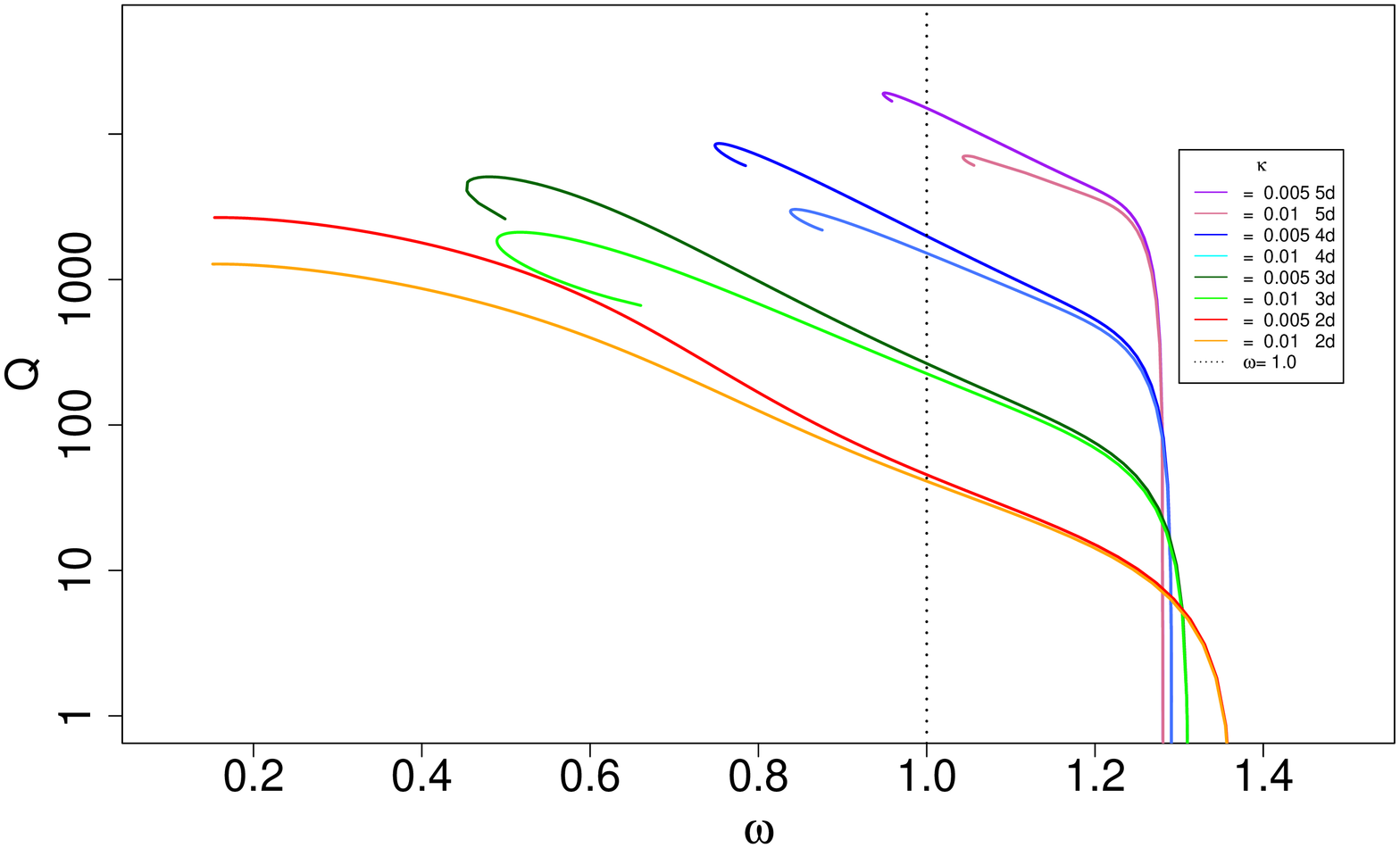}} 
\end{center}
\caption{\label{mq_om_lam} The value of the mass $M$ (left) and the charge $Q$ (right) of the boson stars in dependence
on the frequency $\omega$ in asymptotically AdS space-time $(\Lambda=-0.1$) 
for different values of $d$ and $\kappa$.  }
\end{figure}

As for $\kappa=0$, we find that the mass tends to zero at $\omega\rightarrow \omega_{\rm max}$, while
at $\omega\rightarrow\omega_{\rm min}$ we find the spiraling behaviour typical for boson star solutions.
This is true for all $d\geq 3$. Hence, while boson stars in asymptotically flat space-time with $d\geq 5$ dimensions
can have arbitrarily large masses and charges, their mass and charge are bounded from above in asymptotically
AdS space-time. 

The value of $\omega_{\rm max}$ depends on $\Lambda$ and $d$, but not on $\kappa$. It increases with
decreasing $\Lambda$ and decreasing $d$. Again, this can be explained with the fact that the maximal
value of $\omega$ can be very well approximated by $\omega_{\rm max}=\Delta/\ell$. 

In terms of the holographic interpretation our solutions describe
scalar glueball condensates including backreaction. Our results for the expectation value of the
dual operator on the AdS boundary are given in Fig.\ref{glueballs_backreaction}.

\begin{figure}[h]
\begin{center}
\subfigure[][$<{\rm O}>^{1/\Delta}$ over $\phi(0)$]{\label{condensate_phi0}
\includegraphics[width=5.5cm,angle=270]{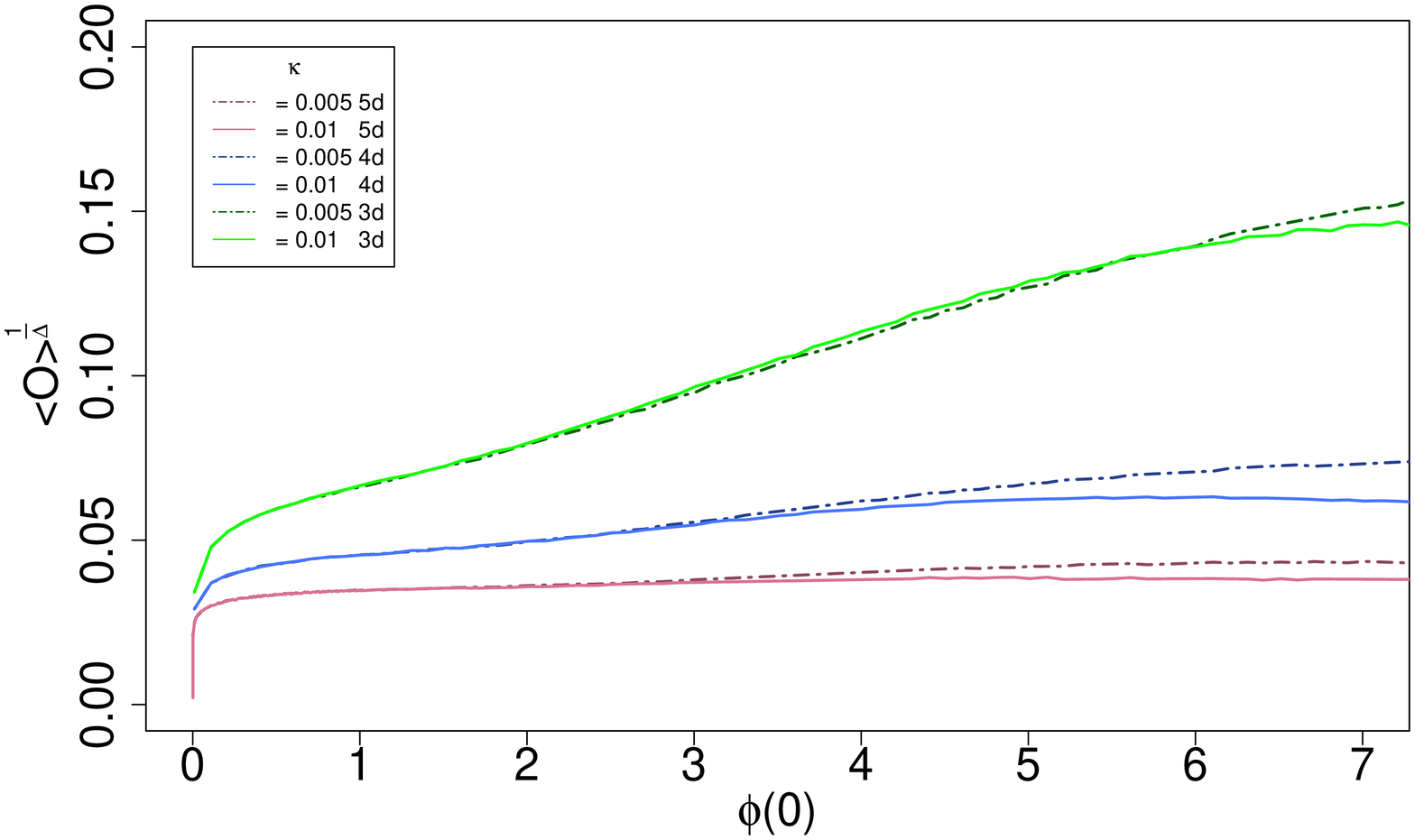}} 
\subfigure[][$<{\rm O}>^{1/\Delta}$ over $M$]{\label{condensate_m}
\includegraphics[width=5.5cm,angle=270]{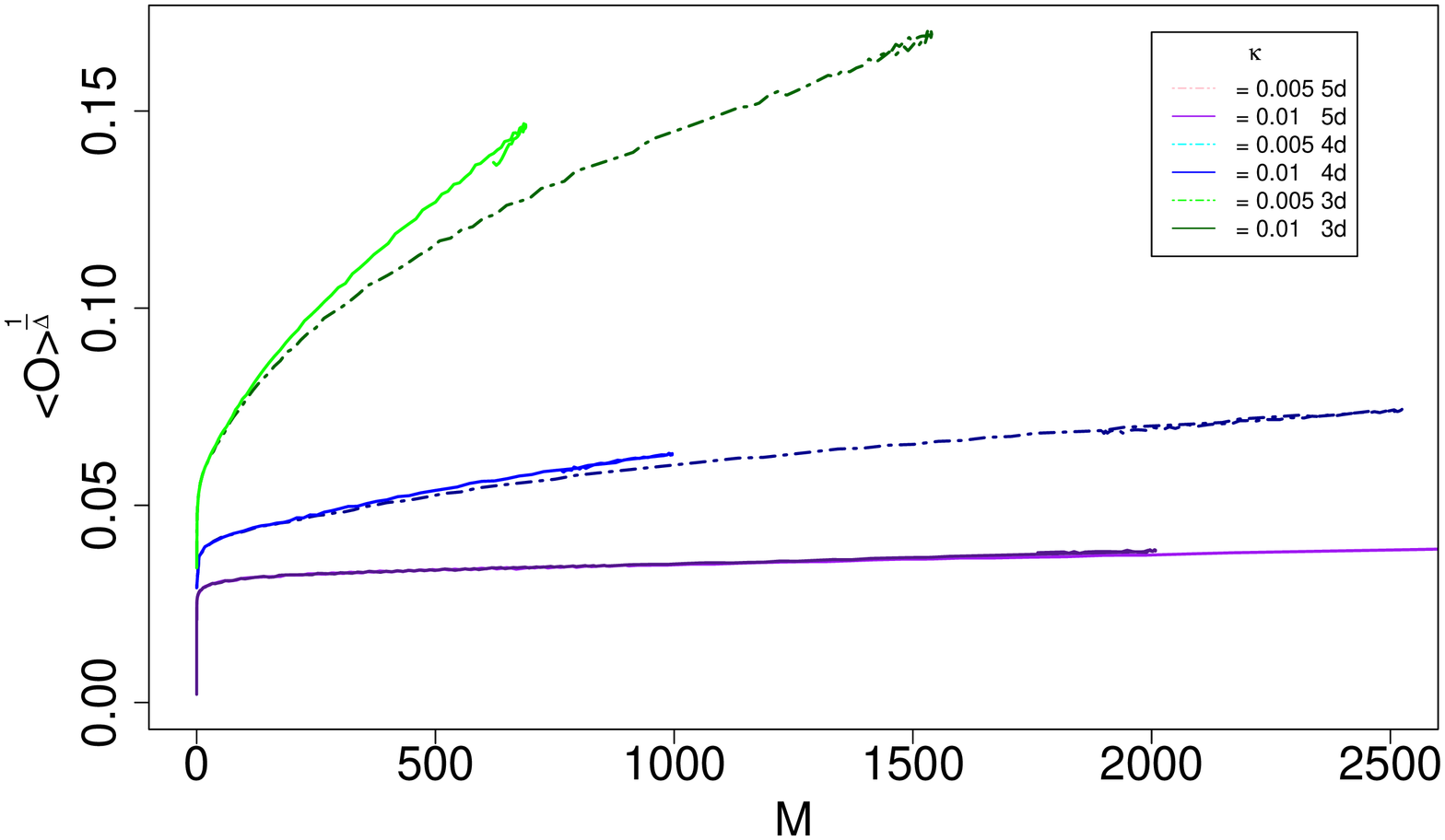}} 
\end{center}
\caption{\label{glueballs_backreaction} We show the expectation value of the dual operator on the AdS boundary
$<{\rm O}>^{1/\Delta}$ corresponding to the value of the condensate of scalar glueballs 
in dependence on $\phi(0)$ (left) and in dependence on $M$ (right) for different values of $\kappa$ and $d$ 
with $\Lambda=-0.1$.   }
\end{figure}

We observe that the value of the condensate at a fixed value of the mass $M$ increases with increasing $\kappa$, i.e. backreaction of the space-time.
This is true for all $d$. Moreover, there exists a maximal possible value of the condensate at the maximal
value of the mass corresponding to intermediate values of $\phi(0)$.

\section{Conclusions and Outlook}
In this paper, we have studied $(d+1)$-dimensional $Q$-balls and boson stars in (asymptotically) 
flat as well as AdS space-time.
We have used an exponential self-interaction potential for the scalar field motivated from
MSSMs. We find that the behaviour of the solutions in asymptotically flat space-time
at the approach of the maximal frequency depends strongly on the number of spatial dimensions $d$ such that
for $d=2,3,4$ boson stars exist only up to a maximal value of the mass and charge, while for $d\geq 5$ the mass
and charge of these objects can become arbitrarily large. However, our numerical results also indicate that
for $d=5$ a mass gap for the solutions exists.
On the other hand, the value of the maximal and minimal frequency does not depend on $d$ in this case.
We show that the existence of boson stars in $(2+1)$-dimensional asymptotically flat space-time depends crucially
on the choice of the scalar potential. For our choice of potential, $(2+1)$-dimensional boson stars always exist.

We also generalize the exact $Q$-ball solution in AdS space-time found in \cite{radu_subagyo} to $d$ dimensions and 
find that
the conclusion about the maximal value of $\omega$ drawn from this solution gives a good approximation
to our numerical results. As such the maximal value of the frequency $\omega$ decreases with decreasing AdS radius
and with $d$. 

We can also draw conclusions about the stability of these objects with respect to the decay into $Q$ free bosons.
While there are stable as well as unstable $Q$-balls in Minkowski space-time for $d\geq 3$, they are
always stable in $d=2$. In AdS space-time $Q$-balls are stable for small values of the charge and
unstable for large values of the charge. Boson stars in asymptotically flat space-time are always stable,
while for asymptotically AdS we find that only for $d=2$ and above a critical $\omega$ the solutions become
stable. For all other $d$ the solutions are stable. 

With view to the holographic interpretation spinning solutions in asymptotically AdS space-time play an important 
role. These solutions has been constructed recently in $d=3$ \cite{radu_subagyo}. 
However, the Ansatz was chosen such that the boundary theory possesses no
rotation and hence describes static glueball condensates. It is surely of interest to generalize this to
find a holographic description of scalar glueballs possessing angular momentum. \\ 
\\
{\bf Acknowledgments} 
We gratefully acknowledge support within the framework of the DFG Research
Training Group 1620 {\it Models of gravity}.

\section{Appendix 1: An exact solution and the value of $\omega_{\rm max}$}
In analogy to the $d=3$ case \cite{radu_subagyo} we find that there exists an exact solution 
when choosing the scalar field potential
of the form $\tilde{U}(\phi)=\mu^2\phi^2 - \lambda \phi^{2k}$ and $\kappa=0$ such that
\begin{equation}
\label{exact}
 \phi(r)=\left(\frac{\mu^2}{\lambda}\frac{\Delta^2 - \ell^2 \omega^2}{(\Delta-d)(\Delta +1)}
\right)^{\Delta/2} \left(1+\frac{r^2}{\ell^2}\right)^{-\Delta/2}
\end{equation}
where $k=1+1/\Delta$.
In the limit $\omega\rightarrow \omega_{\rm max}$ it is known that the function $\phi(r)$ spreads out and becomes
zero everywhere. Hence, we can read of the value of $\omega_{\rm max}$ from (\ref{exact}) which gives
\begin{equation}
 \omega_{\rm max}=\Delta/\ell \ .
\end{equation}

\section{Appendix 2: Existence of gravitating, asymptotically flat boson stars in $2+1$ dimensions}
We follow the calculations in \cite{radu} and employ a scaling argument. For this, we use the
reduced action which reads
\begin{equation}
S=\int\limits_0^{\infty} {\rm d} r\left(\frac{(d-1)}{8\pi G_{d+1}} n' A - r^{d-1} A 
\left(N\phi'^2 - \frac{\omega^2 \phi^2}{N A^2} + U(\phi)\right)\right) \ ,
\end{equation}
where the prime denotes the derivative with respect to $r$.
We now rescale the variable $r\rightarrow \mu r$
and require that $({\rm d}S/{\rm d}\mu)\vert_{\mu=1}=0$. This gives
\begin{eqnarray}
\label{int1}
\int\limits_0^{\infty} {\rm d} r \ r^{d-1} A  \left[(d-2)\left(1-\frac{2\Lambda r^2}
{(d-1)(d-2)}\right)\phi'^2 + d\cdot U(\phi)  
-\left(d - \frac{4(d-1)n}{r^{d-2}}-\frac{2(d-2)\Lambda r^2}
{d(d-1)}\right)\frac{\omega^2\phi^2}{N^2 A^2}\right] = 0
\end{eqnarray}
The equation for $\phi$ (\ref{phi_eq}) can be rewritten as follows
\begin{equation}
\frac{1}{2} \left(r^{d-1} A N (\phi^2)'\right)'= r^{d-1} A 
\left(N(\phi')^2+\frac{1}{2} \frac{\partial U}{\partial \phi}\phi 
-\frac{\omega^2\phi^2}{N A^2}\right)   \ .
\end{equation}
Integrating this with the help of the boundary conditions (\ref{bc1}),(\ref{bc2a}) we find that 
\begin{equation}
\label{int2}
 \int\limits_0^{\infty} {\rm d} r \ r^{d-1} A \left(N(\phi')^2+\frac{1}{2} 
\frac{\partial U}{\partial \phi}\phi 
-\frac{\omega^2\phi^2}{N A^2}\right) = 0  \ .
\end{equation}
Combining with (\ref{int1}) we get
\begin{eqnarray}
\label{full_int}
  \int\limits_0^{\infty} {\rm d} r \ r^{d-1} A \left[ \left(1-\frac{d\cdot n}{r^{d-2}}\right)\phi'^2 +
\frac{d}{2}\left(\frac{1}{2}\frac{\partial U}{\partial \phi}\phi - U(\phi)\right) +
\frac{\omega^2 \phi^2 e^{2\delta}}{N^2}\left(\frac{2\Lambda r^2}{d(d-1)} - \
\frac{(d-2)n}{r^{d-2}}\right) \right] = 0  \ .
\end{eqnarray}
For $\Lambda=0$ and $d=2$ this reads
\begin{equation}
 \int\limits_0^{\infty} {\rm d} r \ r A\left[N\phi'^2 +
\left(\frac{1}{2}\frac{\partial U}{\partial \phi}\phi - U(\phi)\right) \right] = 0 \ .
\end{equation}
Now $N\phi'^2$ is always positive. In \cite{radu} it was argued that for $U(\phi)=m^2\phi^2$ no
gravitating asymptotically flat boson stars in $d=2$ exist. This can be immediately seen when realizing
that in this case the potential term vanishes and the integral can never become zero. However for our case
with $U(\phi)=1-\exp(-\phi^2)$ the potential term gives
\begin{equation}
 \left(\frac{1}{2}\frac{\partial U}{\partial \phi}\phi - U(\phi)\right) = \exp(-\phi^2)\left(\phi^2+1\right) - 1 \ .
\end{equation}
This expression is negative definite (and vanishes for $\phi\equiv 0$) such that we can
have gravitating asymptotically flat boson stars in $d=2$. Note that for the $\phi^6$ self-interaction potential
the negative-definiteness of this expression depends crucially on the choice of potential parameters.
As such, our choice of potential also seems ``more natural'' from this point of view, since
we can have solutions in all possible limits and dimensions of our model without ``fine-tuning'' the potential.

\end{document}